\def\den{\hbox{den}}

\def\tr{\hbox{tr}}
\def\ln{\ell{n}}
\documentstyle[a4,12pt]{article}
\begin{document}
\begin{titlepage} \vspace{0.2in} \begin{flushright}
MITH-96/2 \\ \end{flushright} \vspace*{1.5cm}
\begin{center} {\LARGE \bf  A possible scaling region of chiral fermions on a
lattice\\} \vspace*{0.8cm}
{\bf She-Sheng Xue$^{(a)}$}\\ \vspace*{1cm}
INFN - Section of Milan, Via Celoria 16, Milan, Italy\\ \vspace*{1.8cm}
{\bf   Abstract  \\ } \end{center} \indent
 
We present the details of analyzing an $SU_L(2)\otimes U_R(1)$ chiral theory
with multifermion couplings on a lattice. An existence of a possible scaling
region in the phase space of multifermion couplings for defining the continuum
limit of chiral fermions is advocated. In this scaling region, no spontaneous
symmetry breaking occurs; the ``spectator'' fermion $\psi_R(x)$ is a free mode
and decoupled; doublers are decoupled as massive Dirac fermions consistently
with the $SU_L(2)\otimes U_R(1)$ chiral symmetry, whereas the normal mode of
$\psi_L^i(x)$ is plausibly speculated to be chiral in the continuum limit.
This is not in agreement with the general belief of the definite failure of
theories so constructed. 

\vfill \begin{flushleft}  March, 1996   \vspace*{3cm} \\
\noindent{\rule[-.3cm]{5cm}{.02cm}} \\
\vspace*{0.2cm} \hspace*{0.5cm} ${}^{a)}$ 
E-mail address: xue@milano.infn.it\end{flushleft} \end{titlepage}
 
\section{Introduction}
 
Since the ``no-go'' theorem \cite{nn81} of Nielsen and Ninomiya was demonstrated
in
1981 the problem of chiral fermion ``doubling'' and ``vector-like'' phenomenon
on a lattice still exists if one insists on preserving chiral
symmetry. One of the ideas to get around this ``no-go'' theorem was proposed by
Eichten and Preskill (EP) \cite{ep} ten years ago. The crucial points of this
idea can be briefly described as follows. Multifermion couplings are
introduced such that, in the phase space of strong-couplings, Weyl states
composing three elementary Weyl fermions (three-fermion states) are bound.
Then, these three-fermion states pair up with elementary Weyl fermions to be
Dirac fermions. Such Dirac fermions can be massive without violating chiral
symmetries due to the appropriate quantum numbers and chirality carried by
these three-fermion states. The binding thresholds of such three-fermion states
depend on elementary Weyl modes residing in different regions of the Brillouin
zone. If one assumes that the spontaneous symmetry breaking of the Nambu-Jona
Lasinio (NJL) type does not occur and such binding thresholds separate the weak
coupling symmetric phase from the strong-coupling symmetric phase, there are
two possibilities to realize the continuum limit of chiral fermions in phase
space. One is of crossing over the binding threshold the three-fermion
state of chiral fermions. Another is of a wedge between two thresholds, where
the three-fermion state of chiral fermions has not been formed, provided all
doublers sitting in various edges of the Brillouin zone have been bound to be
massive Dirac fermions and decouple. 

To visualize this idea, EP proposed a model \cite{ep} of multifermion couplings
with $SU(5)$ and $SO(10)$ chiral symmetries and suggested the possible regions
in phase space to define the continuum limit of chiral fermions. However, the
same model of multifermion couplings with $SO(10)$ chiral symmetry was studied
in ref.~\cite{gpr}, it was pointed out that such models of multifermion
couplings fail to give chiral fermions in the continuum limit. The
reasons \footnote{ I thank M.F.L.~Golterman and D.N.~Petcher for discussions on
this subject.} are that an NJL spontaneous symmetry breaking phase separating
the strong-coupling symmetric phase from the weak-coupling symmetric phase, the
right-handed Weyl states do not completely disassociate from the left-handed chiral
fermions and the phase structure of such a model of multifermion couplings is
similar to that of the Smit-Swift (Wilson-Yukawa) model \cite{ss}, which has been
very carefully studied and shown to fail. It is a general
belief \cite{petcher} that the constructions \cite{ep,xue91,xue94} of chiral
fermions on the lattice with multifermion couplings must fail to give chiral
fermions on the basis of a general opinion that multifermion couplings and
Yukawa couplings should be in the same universality class. In fact, this
opinion is indeed correct if one considers multifermion couplings and Yukawa
couplings only for continuous field theory in the sense that two theories have
the same spectrum and relevant operators at the ultra-violate fix
point \cite{shen}. However, it is hard to prove this opinion by showing a
one-to-one correspondence between two phase spaces of multifermion couplings
and Yukawa couplings on a lattice, where they have the exactly same spectra and
relevant operators not only for chiral fermions but also for doublers. Even for
Yukawa couplings, models with different symmetries could be in different
universality classes \cite{gps}. Generally speaking, multifermion couplings possess
more symmetries than Yukawa couplings in a lattice theory. All symmetries of
the Standard Model can possibly be preserved by multifermion couplings, but not
Wilson-Yukawa couplings on a lattice. 

We should not be surprised that a particular model of multifermion couplings
does not work. This does not means that EP's idea is definitely wrong unless there
is another generalized ``no-go'' theorem on interacting theories \cite{ys93} for
a whole range of coupling strength; we will come back to this point in section
6. Actually, Nielsen and Ninomiya gave an interesting comment on EP's idea
based on their intuition of anomalies \cite{nn91}. As a matter of fact, the
phase space of multifermion coupling models of the EP type \cite{ep,gpr} has not
been completely explored. To conclude, we believe that further considerations
of constructing chiral fermions on the lattice with multifermion couplings and
careful studies of the spectrum in each phase of theories so constructed are
necessary. 
 
Note that exploring a possible scaling region of the continuum limit of lattice
(non-gauged) chiral fermions in the phase space of multifermion couplings is
the main goal of this paper \footnote{The short version of this paper is  
published in Phys.~Lett.~B381 (1996) 277.}. This is tightly related to many problems in
particle physics \cite{creutz}. We are not pretending to solve all problems of
lattice chiral gauge theories in this paper. If chiral fermions are gauged, the
important questions concerning correct and consistent features of gauge
bosons and the coupling between gauge bosons and chiral fermions are open,
but beyond the scope of this paper. Other important questions \cite{anomaly} concerning
about gauge anomalies and anomalous global currents (instanton effects and
non-conservation of fermion currents), which EP \cite{ep} suggested to obtain
by explicitly breaking the global symmetries associating to these currents,
will be studied in separate papers. 

In section 2, we present a model of chiral fermions with multifermion couplings
on the lattice and discuss the $\psi_R(x)$ shift-symmetry and its related Ward
identity. The analyses of the weak-coupling phase and the strong-coupling phase
are given in sections 3 and 4. The thresholds and wedges that EP expected are
qualitatively determined and discussed in section 5. An existence of a possible
scaling region for the continuum limit of lattice chiral fermions is advocated
and discussed in section 6. In the last section, we make some remarks on
problems and possible resolutions if this model is chirally gauged. 

\section{Formulation and the $\psi_R$ shift-symmetry} 

Let us consider the following action of chiral fermions with the 
$SU_L(2)\otimes U_R(1)$ global chiral symmetry on the lattice.
\begin{eqnarray}
S&=&S_f+S_1+S_2,\label{action}\\
S_f&=&{1\over 2a}\sum_x\sum_\mu\left(\bar\psi^i_L(x)\gamma_\mu D^\mu_{ij}\psi^j_L(x)+
\bar\psi_R(x)\gamma_\mu\partial^\mu\psi_R(x)\right)\nonumber\\
S_1&=&g_1
\sum_x\bar\psi^i_L(x)\cdot\psi_R(x)\bar\psi_R(x)\cdot\psi_L^i(x)
\nonumber\\
S_2&=&g_2\sum_x \bar\psi^i_L(x)\cdot\left[\Delta\psi_R(x)\right]
\left[\Delta\bar\psi_R(x)\right]\cdot\psi_L^i(x).\nonumber
\end{eqnarray}
In eq.~(\ref{action}), $S_f$ is the naive lattice action of chiral fermions, 
$a$ is the lattice spacing and the $SU_L(2)$ chiral symmetry is actually
local and can easily be gauged
\begin{equation}
\sum_\mu \gamma_\mu D^\mu=\sum_\mu(U_\mu(x)\delta_{x,x+\mu}
-U^\dagger_\mu (x)\delta_{x,x-\mu}),\hskip0.5cm U_\mu(x)\in SU_L(2),
\label{kinetic}
\end{equation}
but we will impose $U_\mu(x)=1$ so that the $SU_L(2)$ is a global symmetry. 
$S_1$ and $S_2$ are two external multifermion 
couplings, where
\begin{eqnarray}
\Delta\psi_R(x)&\equiv&\sum_\mu
\left[ \psi_R(x+\mu)+\psi_R(x-\mu)-2\psi_R(x)\right],\nonumber\\
\Delta\bar\psi_R(x)&\equiv&\sum_\mu
\left[ \bar\psi_R(x+\mu)+\bar\psi_R(x-\mu)-2\bar\psi_R(x)\right].
\label{diff}
\end{eqnarray}
In the action (\ref{action}), $\psi^i_L$ ($i=1,2$) is an $SU_L(2)$ gauged
doublet, $\psi_R$ is an $SU_L(2)$ singlet and both are two-component Weyl
fermions. $\psi_R$ is treated as a ``spectator'' fermion. 
$\psi^i_L$ and $\psi_R$ fields are dimensionful $[a^{-{1\over2}}]$. The first
multifermion coupling $S_1$ in eq.~(\ref{action}) is a dimension-6 operator
relevant both for doublers $p=\tilde p+\pi_A$ and for normal modes $p=\tilde p$
of $\psi^i_L$ and $\psi_R$ fields. Note that all momenta are scaled to be
dimensionless, the physical momentum ($\tilde p$) of normal modes and the
momentum $p=\tilde p+\pi_A$ of doublers are 
\begin{equation}
\tilde p\simeq 0,\hskip1cm p=\tilde p+\pi_A,
\nonumber
\end{equation}
where $\pi_A$ runs over fifteen lattice momenta $\pi_A\not=0$. The second
multifermion coupling $S_2$ in eq.~(\ref{action}) is a dimension-10 operator
relevant only for doublers, but irrelevant for normal modes of $\psi^i_L$
and $\psi_R$. The multifermion couplings $g_1$ and $g_2$ have dimension
$[a^{-2}]$. The $S_1$ is similar to the mass term in lattice QCD and the
second term is similar to the Wilson term. They are quadrilinear in order
to preserve chiral gauge symmetries. 

The action (\ref{action}) has an exact local $SU(2)$ chiral gauge symmetry,
which is the symmetry that the continuum theory (the target
theory) possesses. The global flavour symmetry $U_L(1)\otimes U_R(1)$ is not
explicitly broken in eq.~(\ref{action}). When $g_1=0$, the
action (\ref{action}) possesses a $\psi_R$ shift-symmetry \cite{gp},
i.e., the action is invariant under the transformation:
\begin{equation}
\bar\psi_R(x) \rightarrow \bar\psi_R(x)+\bar\epsilon,\hskip1.5cm
\psi_R(x) \rightarrow \psi_R(x)+\epsilon,
\label{shift}
\end{equation}
where $\epsilon$ is independent of space-time.

To derive the Ward identity associated with this $\psi_R(x)$-shift-symmetry
(\ref{shift}), we consider the generating functional $W(\eta,J)$ and partition
functional $Z(\eta,J)$ of the theory, 
\begin{eqnarray}
W(\eta,J)&=&-\ln Z(\eta,J),
\label{gen}\\
Z(\eta,J)&=&\int_\phi\exp\left(-S+\int_x\left(\bar\psi^i_L\eta_L^i+
\bar\eta_L^i\psi^i_L+\bar\psi_R\eta_R+
\bar\eta_R\psi_R+A_\mu J^\mu\right)\right),
\label{part}\\
\int_\phi &=&\int [d\psi_L^i d\psi_R dA_\mu],\nonumber
\end{eqnarray}
where $A_\mu(x)$ refers to the $SU_L(2)$ gauge field (\ref{kinetic})
defined on the lattice. Then we define the generating functional of
one-particle irreducible vertices (the effective action
$\Gamma(\psi'^i_L,\psi_R',A_\mu')$) as the Legendre transform of $W(\eta, J)$, 
\begin{equation}
\Gamma(\psi'^i_L,\psi'_R, A'_\mu)=W(\eta, J)-\int_x\left(\bar\psi'^i_L\eta_L^i+
\bar\eta_L^i\psi'^i_L+\bar\psi'_R\eta_R+
\bar\eta_R\psi'_R+A'_\mu J^\mu\right),\label{proper}
\end{equation}
with the relations
\begin{eqnarray}
A'_\mu(x)&=&\langle A_\mu (x)\rangle=-{\delta W\over\delta J_\mu (x)},
\nonumber\\
\psi'^i_L(x)&=&\langle\psi^i_L(x)\rangle=-{\delta W\over\delta\bar\eta^i_L(x)
},\hskip1cm
\bar\psi'^i_L(x)=\langle\bar\psi^i_L(x)\rangle={\delta W\over\delta
\eta^i_L(x)},\label{pl}\\
\psi'_R(x)&=&\langle\psi_R(x)\rangle=-{\delta W\over\delta\bar\eta_R(x)},
\hskip1cm
\bar\psi'_R(x)=\langle\bar\psi_R(x)\rangle={\delta W\over\delta\eta_R(x)},
\label{pr}
\end{eqnarray}
in which the fermionic derivatives are left-derivatives, and
\begin{eqnarray}
J_\mu(x)&=&-{\delta\Gamma\over\delta A'_\mu(x)},\nonumber\\
\eta^i_L(x)&=&-{\delta\Gamma\over\delta\bar\psi'^i_L(x)},\hskip1cm
\bar\eta^i_L(x)={\delta\Gamma\over\delta\psi'^i_L(x)},\label{el}\\
\eta_R(x)&=&-{\delta\Gamma\over\delta\bar\psi'_R(x)},\hskip1cm
\bar\eta_R(x)={\delta\Gamma\over\delta\psi'_R(x)}.\label{er}
\end{eqnarray}
In eqs.~(\ref{pl},\ref{pr}), the
$\langle\cdot\cdot\cdot\rangle$ indicates 
\begin{equation}
\langle\cdot\cdot\cdot\rangle={1\over Z}\int_\phi
\left(\cdot\cdot\cdot\right)\exp\left(-S+\int_x\left(\bar\psi^i_L\eta_L^i+
\bar\eta_L^i\psi^i_L+\bar\psi_R\eta_R+
\bar\eta_R\psi_R+A_\mu J^\mu\right)\right),
\label{cdot}
\end{equation}
which is an expectation value with respect to 
the partition functional $Z(\eta, J)$.

Making the parameter $\epsilon$ to be space-time
dependent, and varying the generating functional (\ref{gen}) according to the
transformation rules (\ref{shift}) for arbitrary $\epsilon(x)\not= 0$, we arrive
at 
\begin{equation}
\bar\epsilon(x)\langle {1\over 2a}\gamma_\mu\partial^\mu\psi_R(x)
+g_1\bar\psi^i_L(x)\cdot\psi_R(x)\psi_L^i(x)
+g_2\Delta\left(\bar\psi^i_L(x)\cdot
\Delta\psi_R(x)\psi_L^i(x)\right)+\eta_R(x)\rangle=0.
\label{dw}
\end{equation}
Together with (\ref{er}), the Ward identity in terms of the primed fields 
corresponding to the
$\psi_R$ shift-symmetry of the action (\ref{action}) is given as 
\begin{equation}
{1\over 2a}\gamma_\mu\partial^\mu\psi^\prime_R(x)
+g_1\langle\bar\psi^i_L(x)\!\cdot\psi_R(x)\psi_L^i(x)\rangle
\!+g_2\!\langle\Delta\!\left(\bar\psi^i_L(x)\!\cdot\!
\Delta\psi_R(x)\psi_L^i(x)\right)\rangle-{\delta\Gamma\over\delta\bar
\psi'_R(x)}=0.
\label{w}
\end{equation}
Based on this Ward identity, one can get all one-particle irreducible
vertices containing at least one external $\psi_R$. 

Taking the functional derivative of eq.~(\ref{w}) with respect to $\psi'_R(0)$ 
and then putting external sources $\eta=0$ and $J=0$, we derive 
(see appendix I):
\begin{eqnarray}
&&\left[{1\over 2a}(\gamma_\mu)^{\beta\alpha}\partial^\mu P_R
-g_1\langle\bar\psi^{i\alpha}_L(x)\psi_L^{i\beta}(x)\rangle_\circ
-g_2\Delta\left(\langle\bar\psi^{i\alpha}_L(x)\psi_L^{i\beta}(x)
\rangle_\circ
\Delta\right)\right]\delta(x)\nonumber\\
&&-{\delta^2\Gamma\over\delta\psi'^\alpha_R(0)
\delta\bar\psi'^\beta_R(x)}=0,\label{wf}
\end{eqnarray}
where $\langle\cdot\cdot\cdot\rangle_\circ$ is the expectation value 
(\ref{cdot}) with vanishing external sources $\eta$ and $J_\mu$.
In eqs.~(\ref{pt1},\ref{pt2}) of appendix I, we show
\begin{equation}
\langle\bar\psi^{i\alpha}_L(x)\psi^{i\beta}_L(x)\rangle_\circ=0,\hskip2cm
\Delta\langle\bar\psi^{i\alpha}_L(x)\psi^{i\beta}_L(x)\rangle_\circ=0.
\label{check}
\end{equation}
Thus, the two-point function in eq.~(\ref{wf}) is given as,
\begin{equation}
\int_xe^{-ipx}
{\delta^{(2)}\Gamma\over\delta\psi'_R(x)\delta\bar\psi'_R(0)}={i\over a}
\gamma_\mu\sin (p^\mu a),\label{free}
\end{equation}
which shows that $\psi_R$ does not receive wave-function renormalization.

\section{The weak-coupling region}

Our goal is to seek a possible regime, where an undoubled $SU_L(2)$ chiral
gauged fermion content is exhibited in the continuum limit in the phase space
$(g_1,g_2,g)$, where ``$g$'' is the gauge coupling,  regarded to be a truly
small perturbation $g\rightarrow 0$ at the scale of the continuum limit we
consider. Thus, we impose $g=0$ and $U_\mu(x)=1$ in eq.~(\ref{kinetic}). In the
weak-coupling limit, $g_1\ll 1$ and $g_2\ll 1$ (indicated 1 in Fig.~1), the
action (\ref{action}) defines an $SU_L(2)\otimes U_R(1)$ chiral continuum
theory with a doubled and weakly interacting fermion spectrum that is not the
continuum theory we seek. 

Let us consider the phase of a spontaneous symmetry breaking in the weak-coupling
$g_1,g_2$ limit. Based on the analysis of large-$N_f$ ($N_f$ is an extra
fermion index, e.g.,~color, $N_c$) weak-coupling expansion, we show that the
multifermion couplings in the action (\ref{action}) undergo Nambu-Jona Lasinio
(NJL) spontaneous chiral-symmetry breaking \cite{njl}. In this symmetry breaking
phase indicated 2 in Fig.~1, the $SU_L(2)\otimes U_R(1)$ chiral symmetry is
violated by 
\begin{equation}
{1\over2}\Sigma^i(p)=g_1\int d^4x e^{-ipx}
\langle\bar\psi^i_L(0)\cdot\psi_R(x)\rangle_\circ\not=0.
\label{self}
\end{equation}
Assuming that the symmetry breaking takes place in the direction 1 in the
2-dimensional space of the $SU_L(2)$ chiral symmetry ($\Sigma^1(p)\not=0, 
\Sigma^2(p)=0$), 
one finds the following fermion spectrum that contains a doubled Weyl fermion
$\psi^2_L(x)$ and an undoubled Dirac fermion made up of the Weyl fermions
$\psi^1_L(x)$ and $\psi_R(x)$. The propagators of these fermions can be written
as, 
\begin{eqnarray}
S_{b1}^{-1}(p)&=&{i\over a}\sum_\mu\gamma_\mu\sin p^\mu Z_2(p)P_L
+{i\over a}\sum_\mu\gamma_\mu \sin p^\mu P_R+\Sigma^1(p)\label{sb1}
\\
S_{b2}^{-1}(p)&=&{i\over a}\sum_\mu\gamma_\mu\sin p^\mu Z_2(p)P_L.
\label{sb2}
\end{eqnarray}
The $SU_L(2)\otimes U_R(1)$ chiral symmetry is realized to be $U_L(1)\otimes
U(1)$ with three Goldstone modes and a massive Higgs mode that are not
presented in this report \cite{xue94}.
The fermion self-energy function $\Sigma^1(p)$ (\ref{self}) for $i=1$ 
is given by
\begin{equation}
{1\over2}\Sigma^1(p)=\int d^4xe^{-ipx}
{\delta^{(2)}\Gamma\over\delta\psi'^1_L(x)\delta\bar\psi'_R(0)}.
\label{break}
\end{equation}
The wave-function renormalization $Z_2(p)$ of $\psi_L^i(x)$ field 
is defined as 
\begin{equation}
{i\over a}\gamma_\mu \sin p^\mu\delta_{ij}Z_2(p)=\int d^4xe^{-ipx}
{\delta^{(2)}\Gamma\over\delta\psi'^i_L(x)\delta\bar\psi'^j_L(0)}.
\label{wave}
\end{equation}
The wave-function renormalization of $\psi_R(x)$ field is fixed by
eq.~(\ref{free}).

Based on the Ward identity (\ref{w}) of the $\psi_R$ shift-symmetry, one can 
obtain
an identity for the self-energy function $\Sigma^i(p)$ (\ref{self}).
Performing a functional derivative 
of eq.~(\ref{w}) with respect to $\psi'^i_L(0)$ and
then putting external sources $\eta=0$ and $J=0$, and we obtain (see eqs.~(\ref{der3},
\ref{der4}) in appendix I)
\begin{equation}
g_1\langle\bar\psi^i_L(x)\cdot\psi_R(x)\rangle_\circ\delta(x)
+g_2\langle\Delta\left(\bar\psi^i_L(x)\cdot
\Delta\psi_R(x)\delta(x)\right)\rangle_\circ-
{\delta^2\Gamma\over\delta\psi'^i_L(0)
\delta\bar\psi'_R(x)}=0.
\label{ws1}
\end{equation}
Transforming into momentum space, we obtain
\begin{equation}
{1\over2}\Sigma^i(p)=g_1\langle\bar\psi^i_L(0)\cdot\psi_R(0)\rangle_\circ
+2g_2w(p)\langle\bar\psi^i_L(0)\cdot
\Delta\psi_R(0)\rangle_\circ,
\label{ws2}
\end{equation}
where the well-known Wilson factor \cite{wilson} is
\begin{eqnarray}
w(p)&=&\sum_\mu\left(1-\cos(p_\mu a)\right),\nonumber\\
2w(p)&=&\Delta(p)=\int d^4xe^{-ipx}\Delta(x).
\label{wif}
\end{eqnarray}
As for the four-fermion interaction vertex, analogously, one takes 
functional derivatives of the Ward identity
(\ref{w}) with respect to $\bar\psi'^i_L(0)$, $\psi'^i_L(y)$ and $\psi'_R(z)$
and obtains
\begin{equation}
\int_{xyz}e^{-iyq-ixp-izp'}
{\delta^{(4)}\Gamma\over\delta\psi'^i_L(0)\delta\bar\psi'^i_L(y)\delta\psi'_R(z)
\delta\bar\psi'_R(x)}\!=\! g_1\!+\!4g_2w(p+{q\over 2})w(p'+{q\over 2}),
\label{4p}
\end{equation}
where $p+{q\over 2}$ and $p'+{q\over 2}$ are the momenta of $\psi_R(x)$ field;
$p-{q\over 2}$ and $p'-{q\over 2}$ are the momenta of $\psi^i_L(x)$ field ($q$
is the momentum transfer as shown in Fig.~2.). 
These two identities eqs(\ref{ws2},\ref{4p}) show us two consequences of the 
$\psi_R$ shift-symmetry when $g_1=0$:
(i) the normal modes of $\psi_L^i$ and $\psi_R$ are massless 
\begin{equation}
\Sigma^i(0)=0,\hskip1cm (O(a));
\label{zero}
\end{equation}
(ii) the normal modes of $\psi_R(x)$ and $\psi^i_L(x)$ are free ($O(a^2)$)
from the four-fermion interaction, only the doublers of $\psi_R(x)$ and
$\psi^i_L(x)$ have a non-vanishing four-fermion interacting. We will come back
to these two points in sections 5 and 6. 

Owing to the four-fermion interaction vertex (\ref{4p}), 
the fermion self-energy function $\Sigma^1(p)$ in eqs.~(\ref{self}) and 
(\ref{break}) obeys the NJL gap-equation in the large-$N_f$ weak-coupling
expansion ($N_f\rightarrow\infty$) as shown in Fig.~3,
\begin{equation}
\Sigma^1(p)=4\int_q{\Sigma^1(q)\over\den(q)}\left(\tilde g_1+4\tilde g_2w(p)w(q)
\right)
\label{se}
\end{equation}
where 
\begin{eqnarray}
\int_q&\equiv& \int_{-\pi}^\pi{d^4q\over (2\pi)^4}\nonumber\\
\den(q)&\equiv&
\sum_\rho\sin^2q_\rho +(\Sigma^1(q)a)^2\nonumber\\
\tilde g_1&\equiv& g_1N_fa^2,\hskip1cm \tilde
g_2\equiv g_2N_fa^2.\nonumber
\end{eqnarray}
We adopt the parametrization \cite{gpr}
\begin{equation}
\Sigma^1(p)=\Sigma^1(0)+\tilde g_2 v^1w(p), \hskip1cm 
\Sigma^1(0) =\rho v^1,\label{para}
\end{equation}
where $\rho$ depends only on
couplings $\tilde g_1, \tilde g_2$, and $v^1$ plays a role as the v.e.v.
violating $SU_L(2)\otimes U_R(1)$ chiral symmetry. We can solve the 
gap-equation (\ref{se}) by using this parametrization.
For $v^1=O({1\over a})$, one obtains (see appendix II)
\begin{equation} 
\rho={\tilde g_1\tilde
g_2 I_1\over 1-\tilde g_1 I_\circ}; \hskip1cm \rho={1-4\tilde g_2 I_2\over
4 I_1},\label{rho} 
\end{equation} 
where the functions $I_n(v^1), (n=0,1,2)$, are defined as
\begin{equation}
I_n(v^1)=4\int_q{w^n(q)\over\sum_\rho\sin^2q_\rho +(\Sigma^1(q)a)^2}.
\label{in}
\end{equation}
eq.~(\ref{rho}) leads to a crucial result:
\begin{equation}
\tilde g_1=0,\hskip1cm \rho=0\hskip0.5cm and\hskip0.5cm 
\Sigma^1(0)=0,\label{o}
\end{equation}
this is due to eq.~(\ref{ws2}), resulting from the Ward identity (\ref{w}). 
This means that on the line $g_1$=0, the
normal modes $(p=\tilde p\simeq 0)$ of $\psi^1_L$ and $\psi_R$ are
massless and their 15 doublers ($p=\tilde p+\pi_A$) acquire chiral-variant masses
\begin{equation}
\Sigma^1(p)=\tilde g_2 v^1w(p)
\nonumber
\end{equation}
through the multifermion coupling $g_2$ {\it only}. In this case 
($g_1=0$), the gap-equation is then given by eq.~(\ref{rho}) for $\rho=0$,
\begin{equation}
1-4\tilde g_2 I_2(v^1)=0,\hskip0.3cm i.e.\hskip0.3cm
1=16\tilde g_2\int_q{w^2(q)\over\sum_\rho\sin^2q_\rho +(\tilde g_2v^1w(q)a)^2}.
\label{inte}
\end{equation}
The Wilson factor $w^2(q)$ contained in integral (\ref{inte}) indicates that
only doublers contribute to the NJL gap-equation.

As $v^1\rightarrow 0$, eq.~(\ref{rho}) gives a
critical line $\tilde g^c_1 (\tilde g^c_2)$:
\begin{equation}
\tilde g^c_1={1-4\tilde g_2^c I_2(0)\over4\tilde g_2^cI_1^2(0)
+I_\circ(0)
-4\tilde g_2^cI_\circ(0) I_2(0)},\nonumber
\end{equation} 
characterizing the NJL spontaneous
symmetry breaking. With $I_\circ(0)=2.48, I_1(0)=4I_\circ(0)$ and $
I_2(0)=20I_\circ(0)-4$, the critical points are given by:
\begin{equation}
\tilde g^c_1=0.4,\hskip0.5cm \tilde g^c_2=0;\hskip1.5cm 
\tilde g^c_1=0,\hskip0.5cm \tilde g^c_2=0.0055,\label{wcri}
\end{equation}
as indicated 2 in Fig.~1. These critical values are
sufficiently small even for $N_f=1$. 

As for the wave-function renormalization $Z_2(p)$ in eq.~(\ref{wave}), it
depends on the dynamics of the left-handed Weyl fermion $\psi_L^i$ in this
region. By the large-$N_f$ calculation at weak-couplings (see appendix II and
Figs.4 ,5), we are able to evaluate the wave-function renormalization $Z_2(p)$,
which is given by 
\begin{eqnarray}
Z_2(p)\!&=&\!1\!+\!{2\over N_f}\!\int_{k,q}\!\left(\tilde g_1\!+\!4
\tilde g_2w(p-k)w(k\!+\!{q\over2})\right)^2\!
{\sum_{\mu\nu}\!\gamma_\mu\gamma_\nu
\sin(p\!-\!k)^\mu\sin p^\nu\over\sum_{\lambda\rho}\sin^2(p-k)_\lambda
\sin^2 p_\rho}R(k,\!q)\nonumber\\
R(k,q)&=&{\sum_{\sigma}\sin(k-{q\over2})^\sigma\sin (k+
{q\over2})_\sigma\over\sum_{\sigma\sigma^\prime}
\sin^2(k-{q\over2})^\sigma\sin^2(k+
{q\over2})^{\sigma^\prime}}.
\label{z2}
\end{eqnarray}

This broken phase cannot be a candidate for a real chiral
gauge theory (e.g.,~the Standard Model) for the reasons that $(i)$ $\psi^2_L$ is
doubled (\ref{sb2}); $(ii)$ the spontaneous symmetry breakdown of the
$SU_L(2)$ chiral symmetry is caused by the hard breaking Wilson
term \cite{wilson} (\ref{sb1}) (dimension-5 operator), which must contribute to
the intermediate gauge-boson masses through the perturbative
gauge interaction and disposal of Goldstone modes. The intermediate gauge boson
masses turn out to be $O({1\over a})$. This, however, is
phenomenologically unacceptable. 

\section{The strong-coupling region}

We turn to the strong-coupling region, where $g_1(g_2)$ are sufficiently larger
than a certain critical value $g_1^c(g_2^c)$ (indicated 3 in Fig.~1). 
We can show that the
$\psi^i_L$ and $\psi_R$ in (\ref{action}) are bound to form the
three-fermion states
\begin{equation}
\Psi_R^i={1\over
2a}(\bar\psi_R\cdot\psi^i_L)\psi_R;\hskip1cm\Psi^n_L={1\over 2a}(\bar\psi_L^i
\cdot\psi_R)\psi_L^i.
\label{bound}
\end{equation}
These three-fermion states are Weyl
fermions and respectively pair up with $\bar\psi_R$ and $\bar\psi_L^i$
to be massive, neutral $\Psi_n$ and charged $\Psi_c^i$
Dirac modes, 
\begin{equation}
\Psi^i_c=(\psi_L^i, \Psi^i_R),\hskip1cm\Psi_n=(\Psi_L^n, \psi_R).
\label{di}
\end{equation}
These three-fermion states (\ref{bound}) carry the appropriate quantum
numbers of the chiral group that accommodates $\psi^i_L$ and
$\psi_R$. $\Psi_R^i$ is $SU_L(2)$-covariant and $U_R(1)$ invariant. 
$\Psi_L^n$ is $SU_L(2)$-invariant and $U_R(1)$-covariant. Thus, the spectrum of
the massive composite Dirac fermions $\Psi^i_c$ and $\Psi_n$ is vector-like,
consistent with the $SU_L(2)\otimes U_R(1)$ chiral symmetry. 

In order to study 1PI vertex functions
containing the external legs of three-fermion states (\ref{bound}), we define the
composite ``primed'' fields: the right-handed and charged three-fermion states as
\begin{equation}
\Psi'^i_R\equiv \langle\Psi^i_R\rangle
= {1\over 2a}{\delta^{(3)}
W(\eta)\over\delta\eta_R(x)\delta\bar\eta^i_L(x)\delta\bar\eta_R(x)};
\label{com1}
\end{equation}
and the left-handed and neutral three-fermion states as
\begin{equation}
\Psi'^n_L\equiv \langle\Psi^n_L\rangle={1\over 2a}{\delta^{(3)}
W(\eta)\over\delta\eta^i_L(x)\delta\bar\eta_R(x)\delta\bar\eta^i_L(x)}.
\label{com2}
\end{equation}
Thus, 1PI vertex functions
containing the external legs of three-fermion states (\ref{bound}),
\begin{equation}
{\delta^{(2)}\Gamma\over\delta\Psi'^i_R(x)\bar\psi'^j_L(y)},\hskip0.5cm
{\delta^{(2)}\Gamma\over\delta\Psi'^n_L(x)\bar\psi'_R(y)},\hskip0.5cm
\cdot\cdot\cdot,
\label{1pi}
\end{equation} 
are the truncations of the Green functions 
\begin{eqnarray}
\langle\Psi^i_R(x)\bar\psi^j_L(0)\rangle &=&{1\over 2a}{\delta^{(4)}
W(\eta)\over\delta\eta_R(x)\delta\bar\eta^i_L(x)\delta\bar\eta_R(x)\delta
\eta^j_L(0)},\nonumber\\
\langle\Psi^n_L(x)\bar\psi_R(0)\rangle &=&{1\over 2a} {\delta^{(4)}
W(\eta)\over\delta\eta^i_L(x)\delta\bar\eta_R(x)\delta\bar\eta^i_L(x)\delta
\eta_R(0)},\hskip0.5cm\cdot\cdot\cdot.
\label{green}
\end{eqnarray}
eqs.~(\ref{1pi}) are the most simple couplings between three-fermion states and 
elementary fields.

Now, we study the propagators of composite Dirac fermions (\ref{di}).
As for the neutral composite Dirac fermion, its propagator is given 
as \footnote{I thank 
Y.~Shamir for discussions on these propagators.} 
\begin{equation}
\langle\Psi_n(0)\bar\Psi_n(x)\rangle_\circ=
\langle\Psi^n_L(0)\bar\Psi^n_L(x)\rangle_\circ+\langle\Psi^n_L(0)\bar\psi_R(x)
\rangle_\circ+\langle\psi_R(0)\bar\Psi^n_L(x)\rangle_\circ+\langle\psi_R(0)
\bar\psi_R(x)\rangle_\circ.
\label{dn}
\end{equation}
This propagator can be determined up to a wave
renormalization function by the Ward identity of the $\psi_R$ shift-symmetry.
Taking a functional derivative of the Ward identity eq.~(\ref{w}) with respect to
$\Psi'^n_L(x)$ and then putting external sources $\eta=0$ and $J=0$, we can derive 
\begin{equation}
\int_xe^{ipx}
{\delta^{(2)}\Gamma\over\delta\Psi'^n_L(x)\delta\bar\psi'_R(0)}={1\over2}M(p),
\label{dis}
\end{equation}
where 
\begin{equation}
M(p)=2a(g_1+4g_2w^2(p)).
\label{epmass}
\end{equation}
On the basis of the 1PI vertex functions eqs.~(\ref{free},\ref{dis}), we can 
determine the inverse propagator (\ref{dn}) of the neutral composite
Dirac fermion $\Psi_n(x)$ to be,
\begin{equation}
S_n^{-1}(p)={i\over a}\sum_\mu\gamma_\mu\sin p^\mu Z^n_2(p)P_L
+{i\over a}\sum_\mu\gamma_\mu \sin p^\mu P_R+M(p),
\label{sn}
\end{equation}
where the unknown $Z^n_2(p)$ is a wave-function renormalization for 
$\Psi_L^n(x)$ field.  

The propagator of the charged composite Dirac fermion (\ref{di}) is
\begin{equation}
\langle\Psi^i_c(0)\bar\Psi^j_c(x)\rangle_\circ=
\langle\psi^i_L(0)\bar\psi^j_L(x)\rangle_\circ+\langle\Psi^i_R(0)\bar\psi^j_L(x)
\rangle_\circ+\langle\psi^i_L(0)\bar\Psi^j_R(x)\rangle_\circ+\langle\Psi^i_R(0)
\bar\Psi^j_R(x)\rangle_\circ,
\label{dc}
\end{equation}
which we have to calculate by adopting the strong-coupling expansion.
For the purpose of understanding three-fermion bound states,
we henceforth focus on the region $(g_1\gg 1,
g_2=0)$. We make a rescaling of the fermion fields,
\begin{equation}
\psi_L^i(x)\rightarrow (g_1)^{1\over4}\psi_L^i(x);\hskip1cm
\psi_R(x)\rightarrow (g_1)^{1\over4}\psi_R(x),
\label{rescale}
\end{equation}
and rewrite the action (\ref{action}) and partition function in terms of 
the new fermion fields
\begin{eqnarray}
S_f(x)&=&{1\over 2ag_1^{1\over2}}\sum_\mu\left(\bar\psi^i_L(x)
\gamma_\mu 
\partial^\mu\psi^i_L(x)+
\bar\psi_R(x)\gamma_\mu\partial^\mu\psi_R(x)\right)\label{rfa}\\
S_1(x)&=&\bar\psi^i_L(x)\cdot\psi_R(x)\bar\psi_R(x)\cdot\psi_L^i(x).\label{rs2}
\end{eqnarray}
For the coupling $g_1\rightarrow\infty$, the kinetic
terms $S_f(x)$ can be dropped and we consider this strong-coupling 
limit. With $S_2(x)$ given in eq.~(\ref{rs2}), the integral of $e^{-S_2(x)}$ is 
given (see eq.~(\ref{stronglimit}) with $\Delta(x)=1$ and $g_2\rightarrow g_1$
in appendix III) by
\begin{eqnarray}
Z&=&\Pi_{xi\alpha}\int[d\bar\psi_R^\alpha (x) d\psi_R^\alpha (x)]
[d\bar\psi_L^{i\alpha}(x) d\psi_L^{i\alpha}(x)]\exp\left(-S_1(x)\right)
\nonumber\\
&=&2^{4N},
\label{stronglimit1}
\end{eqnarray}
where ``$N$'' is the number of lattice sites. eq.~(\ref{stronglimit1}) shows a
non-trivial strong-coupling limit. 

We now can perform the strong-coupling expansion in powers of ${1\over g_1}$
about this strong-coupling limit to calculate eq.~(\ref{dc}). To all orders 
in this expansion, the spectrum
of the theory only contains massive states (\ref{dc},\ref{dn}) even though the
$SU_L(2)\otimes U_R(1)$ chiral symmetry is exact. We define two-point
functions in the propagator of the charged Dirac particle (\ref{dc}) to be 
\begin{eqnarray}
S^{ij}_{LL}(x)&\equiv&\langle\psi^i_L(0),\bar\psi^j_L(x)\rangle,\label{sll}\\
S^{ij}_{ML}(x)&\equiv&\langle\psi^i_L(0),[\bar\psi^j_L(x)\cdot\psi_R(x)]
\bar\psi_R(x)\rangle,
\label{sml}\\
S^{ij}_{MM}(x)&\equiv&\langle[\bar\psi_R(0)\cdot\psi^i_L(0)]
\psi_R(0),[\bar\psi^j_L(x)\cdot\psi_R(x)]
\bar\psi_R(x)\rangle.
\label{smm}
\end{eqnarray}
We compute these two-point functions by the strong-coupling expansion in 
powers of $O({1\over g_1})$. Using relations (\ref{p1'},\ref{p3'}) with
$\Delta(x)=1$ and $g_2\rightarrow g_1$ in appendix III, in the lowest 
non-trivial order, we obtain the following recursion relations 
\begin{eqnarray}
S^{ij}_{LL}(x)&=&{1\over g_1}\left({1\over 2a
}\right)^3\sum^\dagger_\mu S^{ij}_{ML}(x+\mu)\gamma_\mu,\label{re1}\\
S^{ij}_{ML}(x)&=&{\delta(x)\delta_{ij}\over 2g_1}
+{1\over g_1}\left({1\over 2a
}\right)\sum^\dagger_\mu S^{ij}_{LL}(x+\mu)\gamma_\mu.
\label{re2}\\
S^{ij}_{MM}(x)&=&{1\over g_1}\left({1\over 2a
}\right)\sum^\dagger_\mu 
\gamma_\mu\gamma_\circ S^{ij\dagger}_{ML}(x+\mu)\gamma_\circ,
\label{re3}
\end{eqnarray}
where for an arbitrary function $f(x)$,
\begin{equation}
\sum_\mu^\dagger f(x)=\sum_\mu \left(f(x+\mu)-f(x-\mu)\right).
\nonumber
\end{equation}
Transforming these two-point functions (\ref{sll},\ref{sml},\ref{smm}) 
into momentum space, ($X=LL,ML,MM$ respectively)
\begin{equation}
S^{ij}_X(p)=\int d^4x e^{-ipx}S^{ij}_X(x),\label{fourier}
\end{equation}
one gets three recursion relations in momentum space
\begin{eqnarray}
S^{ij}_{LL}(p)&=&{1\over g_1}\left({i\over 4a^3
}\right)\sum_\mu \sin p^\mu S^{ij}_{ML}(p)\gamma_\mu,\label{rep1}\\
S^{ij}_{ML}(p)&=&{\delta_{ij}\over 2g_1}
+{i\over g_1a}\sum_\mu\sin p^\mu S^{ij}_{LL}(p)\gamma_\mu.
\label{rep2}\\
S^{ij}_{MM}(p)&=&{1\over g_1}\left({i\over a
}\right)\sum_\mu \sin p^\mu \gamma_\mu\gamma_\circ S^{ij\dagger}_{ML}(p)
\gamma_\circ.
\label{rep3}
\end{eqnarray}
We solve these recursion relations (\ref{rep1},\ref{rep2},
\ref{rep3}) and
obtain
\begin{eqnarray}
S^{ij}_{LL}(p)&=&P_L{\delta_{ij}{i\over2a}\sum_\mu\sin p^\mu\gamma_\mu\over
{1\over a^2}\sum_\mu\sin^2 p_\mu+M^2_1}P_R,\label{sll2}\\
{1\over2a}S^{ij}_{ML}(p)&=&P_L{\delta_{ij}{1\over2}M(p)\over
{1\over a^2}\sum_\mu\sin^2 p_\mu+M^2_1}P_L,\label{slm2}\\
\left({1\over2a}\right)^2S^{ij}_{MM}(p)&=&P_R{\delta_{ij}{i\over2a}\sum_\mu\sin p^\mu\gamma_\mu\over
{1\over a^2}\sum_\mu\sin^2 p_\mu+M^2_1}P_L,\label{smm2}
\end{eqnarray}
where the chiral-invariant mass is
\begin{equation}
M_1=2g_1a.\nonumber
\end{equation}
The second two-point function in eq.~(\ref{dc}) is given by,
\begin{equation}
{1\over2a}\langle[\bar\psi_R(x)\cdot\psi^j_L(x)]
\psi_R(x),\bar\psi^i_L(0)\rangle={1\over2a}\gamma_\circ S_{ML}^{\dagger ij}(x)
\gamma_\circ=P_R{\delta_{ij}{1\over2}M_1\over
{1\over a^2}\sum_\mu\sin^2 p_\mu+M^2_1}P_R.\label{slm3}
\end{equation}
We substitute eqs.~(\ref{sll2},\ref{slm2},\ref{smm2},\ref{slm3}) into 
eq.~(\ref{dc}), in the lowest non-trivial
order of the strong-coupling expansion and obtain the massive propagator of
the charged Dirac fermions $\Psi^i_c$, 
\begin{equation}
S^{ij}_c(p)=\int d^4xe^{-ipx} \langle\Psi^i_c(0)\bar\Psi^j_c(x)\rangle
=\delta_{ij}{{i\over a}\sum_\mu\sin p^\mu\gamma_\mu+M_1\over
{1\over a^2}\sum_\mu\sin^2 p_\mu+M_1^2}.
\label{sc1}
\end{equation}
Analogously, the massive propagator (\ref{dn}) of
the neutral Dirac fermions $\Psi_n$ can be calculated in the same way
\begin{equation}
S_n(p)=\int d^4x e^{-ipx}\langle\Psi_n(0)\bar\Psi_n(x)\rangle
={{i\over a}\sum_\mu\sin p^\mu\gamma_\mu+M_1\over
{1\over a^2}\sum_\mu\sin^2 p_\mu+M_1^2},
\label{sn1}
\end{equation}
which coincides, for $g_2=0$ and $Z^n_2(p)=1$, with eq.~(\ref{sn}) that is
derived by using the Ward identity (\ref{w}) of the $\psi_R(x)$ shift-symmetry.
eqs.~(\ref{sc1},\ref{sn1}) show that the spectrum is vector-like and massive,
consistent with the $SU_L(2)\otimes U_R(1)$ chiral symmetry. In this strong
coupling symmetric phase, all fermion modes including doublers and normal modes
of $\psi^i_L(x)$ and $\psi_R(x)$ are bound to be three-fermion states and
then form massive Dirac fermion states. The spectrum of normal modes and doublers 
is massive and vector-like. This is certainly not what we desire.

\section {Wedges: different thresholds of forming three-fermion states}
 
The three-fermion states (\ref{bound}) are composed of
three elementary Weyl modes through the multifermion couplings $S_1(x)$ and
$S_2(x)$ (\ref{action}). As expected by Eichten and Preskill \cite{ep}, due to
the fact that the multifermion coupling $S_2(x)$ gives different
contributions to the effective value of $g_1$ at large distances for the sixteen
Weyl modes of $\psi_L^i$ and $\psi_R$ in the action (\ref{action}), these
sixteen modes have different thresholds $g^c_1(g^c_2)$ of forming three-fermion
states. In fact, we can explicitly see this point by looking at the
four-fermion 1PI vertex function (\ref{4p}), which is exactly obtained by
the Ward identity (\ref{w}) of the $\psi_R(x)$ shift-symmetry, 
\begin{equation}
\Gamma^{(4)}(p,p',q)=g_1+4g_2w(p+{q\over2})w(p'+{q\over2}),
\label{4p'}
\end{equation}
where $p+{q\over 2}$ and $p'+{q\over 2}$ are momenta of $\psi_R(x)$ field;
$p-{q\over 2}$ and $p'-{q\over 2}$ are momenta of $\psi^i_L(x)$ field ($q$
is the momentum transfer as shown in Fig.~2). In the case of $g_1=0$, the multifermion coupling
associated with the normal modes of $\psi_R(x)$ and $\psi^i_L(x)$ is very 
small ($O(a^2)$). 

These different thresholds $g^c_1(g^c_2)$ can be qualitatively determined by 
the following discussion. We consider 
a complex composite field,
\begin{equation}
{\cal A}^i=\bar\psi_R\cdot\psi^i_L,
\label{comps}
\end{equation}
and its real and imaginary parts are
four composite scalars ($i=1,2$)
\begin{eqnarray}
A_1^i&=&{1\over2}(\bar\psi^i_L\cdot\psi_R+\bar\psi_R\cdot\psi^i_L)
\nonumber\\
A_2^i&=&{i\over2}(\bar\psi^i_L\cdot\psi_R-\bar\psi_R\cdot\psi^i_L).
\label{reals}
\end{eqnarray}
These composite scalars and their propagators are determined by the two-point
function of the theory,
\begin{eqnarray}
G^{ij}(x)&=&\langle{\cal A}^i(0),{\cal A}^{\dagger j}(x)\rangle\nonumber\\
&=&\langle \bar\psi_R(0)\cdot\psi^i_L(0),\bar\psi_R(x)\cdot\psi^j_L(x)
\rangle.
\label{bosonp}
\end{eqnarray}
For simplicity, we put $g_2=0$ and eq.~(\ref{4p'}) becomes 
\begin{equation}
\Gamma^{(4)}(p,p',q)=g_1.
\nonumber
\end{equation}
Adopting the strong-coupling expansion in powers of ${1\over g_1}$
($g_1\gg 1$) and the relation (\ref{p2'})
with $\Delta(x)=1$ and $g_2\rightarrow g_1$ in appendix III, we obtain the following recursion relation
\begin{equation}
G^{ij}(x)={\delta(x)\delta_{ij}\over g_1}
+{1\over g_1}\left({1\over 2a
}\right)^2\sum_{\pm\mu} G^{ij}(x+\mu).
\label{re4}
\end{equation}
Going to momentum space, we have
\begin{equation}
G^{ij}(q)=\int d^4x e^{-iqx} G^{ij}(x),
\nonumber
\end{equation}
where $q$ is the momentum of the composite scalar $
{\cal A}^i=\bar\psi_R\cdot\psi^i_L$. 
The recursion relation (\ref{re4}) in momentum
space is given by
\begin{equation}
G^{ij}(q)={\delta_{ij}\over g_1}+\left(
{1\over 2a^2}\right){1\over g_1}\sum_{\pm\mu}\cos q_\mu G^{ij}(q).
\label{rep4}
\end{equation}
As a result, we find the propagators for these four massive composite scalar 
modes of ${\cal A}^i=\bar\psi_R\cdot\psi^i_L$,
\begin{eqnarray}
G^{ij}(q)&=& 4{\delta_{ij}\over {4\over a^2}\sum_\mu\sin^2{q_\mu\over 2}
+\mu^2};\label{scalar}\\
\mu^2&=& 4\left(g_1-{2\over a^2}\right),
\label{mas}
\end{eqnarray}
which are degenerate owing to the exact $SU_L(2)\otimes U_R(1)$ chiral
symmetry. Thus,
\begin{equation}
\mu^2{\cal A}^i{\cal A}^{i\dagger}
\label{massa}
\end{equation}
effectively gives the mass term of the composite scalar field ${\cal A}^i$ in
the effective Lagrangian. We assume that the 1PI vertex ${\cal A}^j {\cal
A}^{\dagger j}{\cal A}^i{\cal A}^{\dagger i}$ is positive and the energy
of ground states of the theory is bound from the bellow. Then, we can qualitatively
discuss \cite{gpr} the second order phase transition (threshold) from the strong
coupling symmetric phase to the weak-coupling NJL broken phase by examining the
mass term of these composite scalars (\ref{massa}). Spontaneous symmetry
breaking $SU(2)\rightarrow U(1)$ occurs, where $\mu^2>0$ turns to $\mu^2<0$.
eq.~(\ref{mas}) for $\mu^2=0$ gives rise to the critical point: 
\begin{equation}
g_1^ca^2=2,\hskip1cm g_2=0,
\label{gc}
\end{equation}
(as indicated in Fig.~1) where a phase transition takes place between the
NJL symmetry-breaking phase and the strong-coupling symmetric phase. 

As for the case $g_2\not=0$, the second multifermion coupling in
eq.~(\ref{4p'}) has to be taken into account. We should not doubt that the
thresholds $g_1^c(g_2^c)$ depend on the sixteen modes of $\psi_L^i$
and $\psi_R$. In order to see this phenomenon, we can effectively replace the
coupling $g_1$ in eq.~(\ref{mas}) by the coupling (\ref{4p'}) involving $g_2$.
As a result, the thresholds of binding three-fermion states
can be qualitatively determined by 
\begin{equation}
\mu^2= 4\left(g_1+4g_2w(p+{q\over2})w(p'+{q\over2})-{2\over a^2}\right)=0.
\label{mas2}
\end{equation}
Let us first consider the multifermion couplings of each mode ``$p$'' of the
$\psi_L^i$ and $\psi_R$, namely, we set $p=p', q\ll 1$ in the four-point vertex
(\ref{4p'}). We obtain
\begin{equation}
\mu^2= 4\left(g_1+4g_2w^2(p)-{2\over a^2}\right).\label{threshold2'}
\end{equation}
Thus, $\mu^2=0$ gives rise to the critical
lines (thresholds):
\begin{equation}
g_1^ca^2=2, g_2=0;\hskip0.5cm g_1=0, a^2g_2^{c,b}=0.008,
\label{threshold1}
\end{equation}
where the first binding threshold of the doubler $p=(\pi,\pi,\pi,\pi)$ is
located, 
and 
\begin{equation}
g_1^ca^2=2, g_2=0;\hskip0.5cm g_1=0, a^2g_2^{c,a}=0.124,
\label{thresholdl}
\end{equation}
where the last binding threshold of the doublers $p=(\pi,0,0,0)$ is located. In
between
(indicated 4 in Fig.~1) there are the binding thresholds of the doublers
$p=(\pi,\pi,0,0)$ and $p=(\pi,\pi,\pi,0)$ in eq.~(\ref{threshold2'}), and the binding thresholds of the
different doublers $p\not=p'$ in eq.~(\ref{mas2}). Above $g_1^{c,a}$ {\it
all doublers are supposed to be bound}. As for the
normal modes ($\tilde p$) of $\psi^i_L$ and $\psi_R$, when $g_1\ll 1$, 
the multifermion
coupling (\ref{4p'}), $\Gamma^{(4)}=g_1+4g_2w^2(\tilde p)$, is no
longer strong enough to form the three-fermion states
$(\bar\psi^i_L\cdot\psi_R)\psi^i_L$ and $(\bar\psi_R\cdot\psi^i_L)\psi_R$ unless
$a^2g_2 \rightarrow\infty$. It is conceivable that the threshold for binding
normal modes, which is given by eq.~(\ref{threshold2'}) for $p=\tilde p$,
\begin{equation}
g_1+ag_2O(\tilde p^2)-{2\over a^2}=0,
\nonumber
\end{equation}
analytically continues to the limit 
\begin{equation}
g_2^{c,\infty}\rightarrow \infty,\hskip0.5cm g_1\rightarrow 0,
\label{thresholdn}
\end{equation}
as indicated in Fig.~1. Certainly, these thresholds cannot be regarded as
quantitative results obtained by precise computations.

Thus, as expected in ref.~\cite{ep}, several wedges open up as $g_1, g_2$
increase in the NJL phase (indicated 5 in Fig.~1), in between the critical lines
along which three-fermion states of normal modes and doublers of $\psi^i_L$
and $\psi_R$ respectively approach their thresholds. In the initial part of the
NJL phase (the deep NJL broken phase), the normal modes and doublers of the
$\psi^i_L$ and $\psi_R$ undergo the NJL phenomenon and contribute to
eqs.~(\ref{sb1},\ref{sb2}), as discussed in section 2. As $g_1,g_2$ increase, all
these modes, one after another, gradually disassociate from the NJL phenomenon
and no longer contribute to eqs.~(\ref{sb1},\ref{sb2}). Instead, they turn to
associate with the EP phenomenon and contribute to eqs.~(\ref{sc1},\ref{sn1}). 
The first and last doublers of $\psi^i_L$ and $\psi_R$ making this
transition are $p=(\pi,\pi,\pi,\pi)$ (\ref{threshold1}) and $p=(\pi,0,0,0)$
(\ref{thresholdl}) respectively. At the end of this sequence, normal modes
($p=\tilde p)$ make this transition (\ref{thresholdn}), due to the fact that
they possess the different effective multifermion coupling (\ref{4p'}). There
are two possibilities that one might expect to have the continuum limit of chiral
fermions defined either on one of these thresholds or within one of these wedges. 

Had these thresholds separated the two symmetric phases,
(strong-couplings and the weak-coupling symmetric phases) we would have found a
threshold over which an EP phase transition takes place, namely, the massless 
normal modes of the
$\psi_L^i$ and $\psi_R$ are becoming massive, while the doublers
of $\psi_L^i$ and $\psi_R$ have acquired chiral-invariant masses and
decoupled (\ref{sc1},\ref{sn1}). We would define a continuum
theory of massless free chiral fermions \cite{ep} on such a threshold. However,
this is not the real case \cite{gpr} for $g_2=0$, as has been seen in
eq.~(\ref{mas}), $\mu^2>0$ turning to $\mu^2<0$ at $g_1=g^c_1=2$ indicates a
phase transition between the strong-coupling symmetric phase and the spontaneous
symmetry breaking phase, which separates the strong-coupling and weak
coupling symmetric phases. Namely, there is no EP phase transition taking
place over any one of those thresholds. This can be clearly seen (as shown in 
Fig.~1) by comparison of the thresholds of three-fermion states 
(\ref{gc},\ref{threshold1},\ref{thresholdl}) 
and the NJL phase transition (\ref{wcri}).

Had one of these wedges contained a spectrum in which all doublers of $\psi_L^i$
and $\psi_R$ decouple as Dirac fermions by acquiring chiral-invariant masses
(\ref{sc1},\ref{sn1}) and the normal modes of $\psi_L^i$ and $\psi_R$ remain
massless and free, and within that wedge we would have obtained a continuous
theory of massless free chiral fermions \cite{ep}. However, this seems not 
really the case for the reason that all these wedges from the deep NJL broken phase
to the deep EP symmetric phase are continuously connected. These thresholds and
wedges qualitatively determined cannot be considered to be very distinct for
the spectrum of the theory, since there may be the coexistence of less tightly
bound three-fermion states of doublers and unbound doublers, the possible
strong fluctuations of these three-fermion states and mixing between 
different modes.

Even within the last wedge, as indicated 5 in Fig.~1, if we assume that there is
a such region where the couplings $g_1$ and $g_2$ are sufficiently larger than
the threshold (\ref{thresholdl}) where {\it all} doublers are supposed to be
{\it strongly bound\/} \footnote{Here, by the notion of {\it strongly bound}, 
we mean that all doublers are completely and tightly bound into three-fermion 
states and no doublers are left in the spectrum.}, we should have the undoubled low-energy spectrum that
involves only the normal modes of $\psi^i_L$ and $\psi_R$. However, because
of the multifermion coupling $g_1\not= 0$, these normal modes of $\psi^i_L$ and
$\psi_R$ still remain in the NJL broken phase, the $SU_L(2)$ chiral symmetry is
still violated by $\Sigma^1(0)=\rho v^1$ (\ref{para}), to which only normal
modes contribute. The propagators of the normal
modes in this wedge should be the same as eqs.~(\ref{sb1},\ref{sb2}) for
$p=\tilde p$. Due to the fact that chiral gauge symmetry is broken by the
NJL phenomenon associated with the normal modes, the propagator of gauge bosons
will not have the correct properties. Furthermore, when $g_1\not= 0$, the normal 
mode of $\psi_R$ is not
guaranteed to completely decouple from that of $\psi^i_L$. 

So far, we have almost no possibility to find a distinct threshold for the
second-order phase transition to define the continuum limit for chiral fermions
and a distinctly chiral-symmetric region in the phase space where a desired
spectrum of chiral fermions exists. Nevertheless, a possible resolution of
this undesirable situation could probably take place in a particular symmetric
phase that is a segment in the phase diagram, where the doublers of $\psi^i_L$
and $\psi_R$ have formed three-fermion states $(\bar\psi_R\cdot\psi^i_L)\psi_R$ and
$(\bar\psi^i_L\cdot\psi_R)\psi^i_L$ via the EP phenomenon, while the normal
modes of $\psi^i_L$ and $\psi_R$ have neither formed such bound states yet
and nor are they associated with the NJL-phenomenon. If this case happened,
we might find a scaling region of the continuum limit of lattice chiral
fermions. 

\section {A possible scaling region of chiral fermions}

The possible scaling region of the continuum limit of lattice chiral fermion
can probably be found once we go onto the 
line where $g_1=0$ and $g_2$ is in a certain segment A 
indicated in Fig.~1. Assuming that above $g_2^{c,a}$ all doublers are
{\it strongly bound}, above $g_2^{c,\infty}$ all modes are {\it strongly bound}
and $g_2^{c,\infty}>g_2^{c,a}\gg 1$, we give the definition of this peculiar
segment {\it A}:
\begin{equation}
A=\Big[g_1=0, g_2^{c,a}<g_2<g_2^{c,\infty}\Big].
\label{segment}
\end{equation}
In order to show this segment could be a possible candidate for the scaling 
region of the
continuum limit of lattice chiral fermions, we are bound to demonstrate the
following properties of the theory in this segment: 
\begin{itemize}
\begin{enumerate}
\item the normal mode of $\psi_R$ is a free mode and decoupled; 

\item no spontaneous chiral symmetry breaking occurs ;

\item all doublers are {\it strongly bound} to be massive
Dirac fermions and decoupled consistently with chiral symmetry;

\item the normal modes of $\psi^i_L(x)$ and $\psi_R(x)$ have not yet been
bound to the three-fermion state $(\bar\psi_R\cdot\psi^i_L)\psi_R$, an
undoubled chiral mode of $\psi^i_L(x)$ exists in the low-energy spectrum. 

\end{enumerate}
\end{itemize}

We have already discussed the first property in section 2. In the case of $g_1=0$, the
normal mode of $\psi_R(x)$ is massless and does not receive the wave-function
renormalization,
\begin{equation}
S^{-1}_{RR}=i\gamma_\mu\tilde p^\mu.
\nonumber
\end{equation}
Furthermore, in section 4 based on the Ward identity (\ref{w})
associated with the $\psi_R(x)$ shift-symmetry, we have determined the
propagator of the neutral Dirac particle (\ref{sn}), which is made up of the
$\psi_R(x)$ and its left-handed composite partner $\Psi^n_L(x)$ (\ref{bound}),
up to a wave function renormalization for $\Psi^n_L(x)$ field. 

As for the other properties, since $g_1=0, g_2\gg 1$ in this segment, we can adopt
the strong-coupling expansion in powers of ${1\over g_2}$ to calculate 1PI
functions that describe those properties of the theory. Thus, we make a rescaling
of the fermion fields, 
\begin{equation}
\psi_L^i(x)\rightarrow (g_2)^{1\over4}\psi_L^i(x);\hskip1cm
\psi_R(x)\rightarrow (g_2)^{1\over4}\psi_R(x),
\label{rescale1}
\end{equation}
and rewrite the action (\ref{action}) 
and partition function in terms of new fermion fields
\begin{eqnarray}
S_f(x)&=&{1\over 2ag_2^{1\over2}}\sum_\mu\left(\bar\psi^i_L(x)
\gamma_\mu \partial^\mu\psi^i_L(x)+
\bar\psi_R(x)\gamma_\mu\partial^\mu\psi_R(x)\right)\label{rfa1}\\
S_2(x)&=&\bar\psi^i_L(x)\cdot\left[\Delta\psi_R(x)\right]
\left[\Delta\bar\psi_R(x)\right]\cdot\psi_L^i(x).\label{rs21}
\end{eqnarray}
For the coupling $g_2\rightarrow\infty$, the kinetic terms $S_f(x)$
can be dropped and we consider this strong-coupling limit. With $S_2(x)$
given in eq.~(\ref{rs21}), the integral of $e^{-S_2(x)}$ can be computed 
(see the beginning of appendix III) 
\begin{eqnarray} 
Z&=&\Pi_{xi\alpha}\int[d\bar\psi_R^\alpha (x) d\psi_R^\alpha (x)]
[d\bar\psi_L^{i\alpha}(x) d\psi_L^{i\alpha}(x)]\exp\left(-S_2(x)\right)
\nonumber\\
&=&2^{4N}\left(\det\Delta^2(x)\right)^4,
\label{stronglimit}
\end{eqnarray}
where the determent is taken only over the lattice space-time and $N$
is the number of lattice sites. For the non-zero eigenvalues of the operator
$\Delta^2(x)$, eq.~(\ref{stronglimit}) shows an existence of the sensible 
strong-coupling limit. However, as for the zero eigenvalue of the operator
$\Delta^2(x)$, this strong-coupling limit is trivial and should not be analytic and the 
strong-coupling expansion in powers of ${1\over g_2}$ breaks down. 

We consider the second property that this segment is entirely symmetric. The
vanishing of the 1PI self-energy function $\Sigma(p)$ (\ref{self}) indicates no
spontaneous symmetry breaking of the theory. In sections 2 and 3, on the basis
of both the Ward identity (\ref{ws2}) of the $\psi_R$ shift-symmetry and the
explicit computation in the large-$N_f$ weak-coupling expansion (\ref{o}), we
show that this 1PI self-energy function $\Sigma(p)$ vanishes at  
zero momentum, provided $g_1=0$,
\begin{equation}
\Sigma(p)=0, \hskip2cm p=0.
\label{p=0}
\end{equation}
Obviously, we need to show that this 1PI self-energy function $\Sigma(p)$
vanishes for $p\not=0$ in this segment as well. For this purpose, we have to
calculate the following two-point functions 
\begin{eqnarray}
S^j_{RL}(x)&\equiv&\langle\psi_R(0),\bar\psi^j_L(x)\rangle,\label{srl}\\
S^j_{MR}(x)&\equiv&\langle\psi_R(0),[\bar\psi^j_L(x)\cdot\psi_R(x)]
\bar\psi_R(x)\rangle.
\label{smr}
\end{eqnarray}
In appendix III, using the strong-coupling expansion in powers $O({1\over
g_2})$ above the non-trivial strong-coupling limit (\ref{stronglimit})
for the non-zero eigenvalues of $\Delta^2(x)$, we obtain the relations
(\ref{p1'},\ref{p3'}). Armed with these relations (\ref{p1'},\ref{p3'}), in the
lowest non-trivial order, one gets the following recursion relations 
\begin{eqnarray}
S^j_{RL}(x)&=&{1\over g_2\Delta^2(x)}\left({1\over 2a
}\right)^3\sum^\dagger_\mu S^j_{MR}(x+\mu)\gamma_\mu,\label{re5}\\
S^j_{MR}(x)&=&{1\over g_2\Delta^2(x)}\left({1\over 2a
}\right)\sum^\dagger_\mu S^j_{RL}(x+\mu)\gamma_\mu.
\label{re6}
\end{eqnarray}
The Fourier transformations of these two-point functions (\ref{srl},\ref{smr}) 
into momentum space are
\begin{eqnarray}
S^j_{RL}(p)&=&\int d^4x e^{-ipx}S^j_{RL}(x)\nonumber\\
S^j_{MR}(p)&=&\int d^4x e^{-ipx}S^j_{MR}(x).\nonumber
\end{eqnarray}
For $p\not=0$ and the non-zero eigenvalue $\Delta(p)=2w(p)\not=0$ 
(\ref{wilson}), one gets 
two recursion relations in momentum space,
\begin{eqnarray}
S^j_{RL}(p)&=&{1\over 4g_2w^2(p)}\left({i\over 4a^3
}\right)\sum_\mu \sin p^\mu S^j_{MR}(p)\gamma_\mu,\label{rep5}\\
S^j_{MR}(p)&=&{1\over gw^2(p)}\left({i\over 4a
}\right)\sum_\mu\sin p^\mu S^j_{RL}(p)\gamma_\mu.
\label{rep6}
\end{eqnarray}
The solution to these recursion relations is
\begin{equation}
\left((8ag_2w^2(p))^2+{1\over a^2}\sum_\mu\sin^2p_\mu\right)S_{RL}^j(p)=0.
\label{sigmas}
\end{equation}
For $p\not=0$, clearly we must have
\begin{equation}
S^j_{RL}(p)=0,\hskip0.5cm
S^j_{MR}(p)=0.
\end{equation}
Similarly, we can prove the vanishing of the following two point
functions
\begin{eqnarray}
&&\langle[\bar\psi^i_L(0)\cdot\psi_R(0)]\psi^i_L(0),\bar\psi^j_L(x)\rangle,
\nonumber\\
&&\langle[\bar\psi^i_L(0)\cdot\psi_R(0)]\psi^i_L(0),[\bar\psi^j_L(x)
\cdot\psi_R(x)]\bar\psi_R(x)\rangle.
\nonumber
\end{eqnarray}
This demonstration can be straightforwardly generalized to show the vanishing of
all n-point Green functions that are not $SU_L(2)\otimes U_R(1)$ chirally
symmetric. As a consequence, the segment is entirely symmetric and no spontaneous
symmetry breaking takes place. 

Now we turn to discussions of the third properties that all doublers 
are decoupled as massive Dirac fermions consistently
with chiral symmetry.
Analogously to the case $g_1\gg 1, g_2=0$ in section 4, we need to compute the
two-point functions (\ref{sll},\ref{sml},\ref{smm}) in the propagator of
the charged Dirac fermion (\ref{dc}). Performing a strong-coupling expansion in
powers of ${1\over g_2}$ above the non-trivial strong-coupling limit
(\ref{stronglimit}) for non-zero eigenvalues of $\Delta^2(x)$, we compute these two-point functions.
Armed with the relations (\ref{p1'}), (\ref{p3'}) and (\ref{p4}) that are 
obtained in appendix III, in the lowest non-trivial order, we obtain
following recursion relations 
\begin{eqnarray}
S^{ij}_{LL}(x)&=&{1\over g_2\Delta^2(x)}\left({1\over 2a
}\right)^3\sum^\dagger_\mu S^{ij}_{ML}(x+\mu)\gamma_\mu,\label{re11}\\
S^{ij}_{ML}(x)&=&{\delta(x)\delta_{ij}\over 2g_2\Delta^2(x)}
+{1\over g_2\Delta^2(x)}\left({1\over 2a
}\right)\sum^\dagger_\mu S^{ij}_{LL}(x+\mu)\gamma_\mu.
\label{re21}\\
S^{ij}_{MM}(x)&=&{1\over g_2\Delta^2(x)}\left({1\over 2a
}\right)\sum^\dagger_\mu 
\gamma_\mu\gamma_\circ S^{ij\dagger}_{ML}(x+\mu)\gamma_\circ.
\label{re31}
\end{eqnarray}
Upon the non-trivial strong-coupling limit (\ref{stronglimit}) of $p\not=0$
and $\Delta(p)=2w^2(p)\not=0$, Fourier transformation (\ref{fourier}) casts 
three recursion relations in momentum space, 
\begin{eqnarray}
S^{ij}_{LL}(p)&=&{1\over 4g_2w^2(p)}\left({i\over 4a^3
}\right)\sum_\mu \sin p^\mu S^{ij}_{ML}(p)\gamma_\mu,\label{rep11}\\
S^{ij}_{ML}(p)&=&{\delta_{ij}\over 8g_2w^2(p)}
+{i\over 4g_2w^2(p)a}\sum_\mu\sin p^\mu S^{ij}_{LL}(p)\gamma_\mu.
\label{rep21}\\
S^{ij}_{MM}(p)&=&{1\over 4g_2w^2(p)}\left({i\over a
}\right)\sum_\mu \sin p^\mu \gamma_\mu\gamma_\circ S^{ij\dagger}_{ML}(p)
\gamma_\circ.
\label{rep31}
\end{eqnarray}
We solve these recursion relations (\ref{rep11},\ref{rep21},
\ref{rep31}) and obtain
\begin{eqnarray}
S^{ij}_{LL}(p)&=&P_L{\delta_{ij}{i\over2a}\sum_\mu\sin p^\mu\gamma_\mu\over
{1\over a^2}\sum_\mu\sin^2 p_\mu+M_2^2(p)}P_R,\label{sll21}\\
{1\over2a}S^{ij}_{ML}(p)&=&P_L{\delta_{ij}{1\over2}M_2(p)\over
{1\over a^2}\sum_\mu\sin^2 p_\mu+M_2^2(p)}P_L,\label{slm21}\\
\left({1\over2a}\right)^2S^{ij}_{MM}(p)&=&P_R{\delta_{ij}{i\over2a}\sum_\mu\sin p^\mu\gamma_\mu\over
{1\over a^2}\sum_\mu\sin^2 p_\mu+M_2^2(p)}P_L,\label{smm21}
\end{eqnarray}
where the chiral-invariant mass is given as
\begin{equation}
M_2(p)=8g_2aw^2(p),\hskip1cm p\not=0\nonumber
\end{equation}
The second two-point function in eq.~(\ref{dc}) is given by,
\begin{equation}
{1\over2a}\langle[\bar\psi_R(x)\cdot\psi^j_L(x)]
\psi_R(x),\bar\psi^i_L(0)\rangle={1\over2a}\gamma_\circ S_{ML}^{\dagger ij}(x)
\gamma_\circ=P_R{\delta_{ij}{1\over2}M_2(p)\over
{1\over a^2}\sum_\mu\sin^2 p_\mu+M_2^2(p)}P_R.\label{slm31}
\end{equation}
We substitute eqs.~(\ref{sll21},\ref{slm21},\ref{smm21},\ref{slm31}) into
eq.~(\ref{dc}) and obtain the propagator of the charged Dirac doublers $\Psi^i_c$
($p\not=0$), 
\begin{equation}
S^{ij}_c(p)=\int d^4x e^{-ipx}\langle\Psi^i_c(0)\bar\Psi^j_c(x)\rangle
=\delta_{ij}{{i\over a}\sum_\mu\sin p^\mu\gamma_\mu+M_2(p)\over
{1\over a^2}\sum_\mu\sin^2 p_\mu+M_2^2(p)}.
\label{sc11}
\end{equation}
Since $w^2(p)\not=0$ and $M_2(p)\not=0$
for doublers $p=\tilde p+\pi_A$, this shows that all $SU_L(2)$ charged doublers
are decoupled as massive Dirac fermions. The massive spectrum for
these doublers turns out to be $SU(2)$-QCD vector-like, however, is in
consistent agreement with the $SU_L(2)\otimes U_R(1)$ chiral symmetry. 
 
Analogously, in the lowest non-trivial order of the strong-coupling expansion
in powers of ${1\over g_2}$, one gets for $p\not=0$ and $\Delta(p)=2w(p)\not
=0$
\begin{eqnarray}
S_{RR}(p)&=&P_L{{i\over2a}\sum_\mu\sin p^\mu\gamma_\mu\over
{1\over a^2}\sum_\mu\sin^2 p_\mu+M_2^2(p)}P_R,\label{srr21}\\
{1\over2a}S_{MR}(p)&=&P_L{{1\over2}M_2(p)\over
{1\over a^2}\sum_\mu\sin^2 p_\mu+M_2^2(p)}P_L,\label{srm21}\\
\left({1\over2a}\right)^2S_{MM}(p)&=&P_R{{i\over2a}\sum_\mu\sin p^\mu
\gamma_\mu\over
{1\over a^2}\sum_\mu\sin^2 p_\mu+M_2^2(p)}P_L,\label{srmm21}
\end{eqnarray}
where $S_{RR}(p), S_{MR}(p)$ and $S_{MM}(p)$ are the Fourier transformation of
the two-point functions in the neutral Dirac fermion propagator (\ref{dn}).
As a result, the propagator (\ref{dn}) of
the neutral Dirac doublers $\Psi_n$ ($p\not=0$) is given by
\begin{equation}
S_n(p)=\int d^4xe^{-ipx} \langle\Psi_n(0)\bar\Psi_n(x)\rangle
={{i\over a}\sum_\mu\sin p^\mu\gamma_\mu+M_2(p)\over
{1\over a^2}\sum_\mu\sin^2 p_\mu+M_2^2(p)},
\label{sn11}
\end{equation}
which coincides, for $g_1=0$ and $Z^n_2(p)=1 (p\not=0)$, with eq.~(\ref{sn}),
which is 
derived by using the Ward identity (\ref{w}) of the $\psi_R(x)$ shift-symmetry. 
eq.~(\ref{sn11}) shows that the spectrum for doublers of the neutral Dirac 
fermion is vector-like and massive, consistent with the $SU_L(2)\otimes
U_R(1)$ chiral symmetry. 

Up to now, we are left with the last, but most important property that the normal modes of the
$\psi^i_L(x)$ and $\psi_R(x)$ are massless and chiral in this segment. It
is most difficult to have a convincing proof of this property for the time being, since
for these normal modes ($p=\tilde p=0, \Delta(\tilde p)=0$ in
eq.~(\ref{stronglimit})), a sensible non-trivial strong-coupling limit does not
exist and the strong-coupling expansion in powers of ${1\over g_2}$
fails to converge analytically. We are actually not able to compute the spectra
(propagators) of these normal modes to see whether or not they are chiral in
this segment. As a consequence of the interacting action (\ref{action})
presented in this paper being local, the effective action (inverse propagator)
that is bilinear in terms of interpolating fields should be local and
analytical in the whole Brillouin zone (this statement has not received a
complete proof \cite{xy} for the entire range of interacting strength), thus, the
``no-go'' theorem of Nielsen and Ninomiya is still applicable to this
case \cite{ys93}. Based on this observation, one might argue that the massless
spectrum of normal modes should still be vector-like by the analytic
continuation of the charged and neutral Dirac propagators (\ref{sc11}) and
(\ref{sn11}) from $p\not=0$ to $p=\tilde p=0$. This argument is indeed correct
in the phase diagram where $g_1\gg 1$ since we have a sensible strong-coupling 
limit, as we have shown in section 4. While, in
the segment A ($g_2\gg 1$) that we speculate to be the scaling region of
lattice chiral fermions, we are still lacking knowledge and evidence
concerning the analyticity of the strong-coupling limit (\ref{stronglimit})
and the spectrum (\ref{sc11},\ref{sn11}) around the point $p\simeq 0$. The 
problem that confronts us is how the three-fermion states that are composed of
the normal
modes of $\psi_L^i$ and $\psi_R$ disappear. If these three-fermion states
were resonant states, the vacuum would be unstable.
The only loophole would appear if the propagators of interpolating fields (three-fermion
states) vanished at $p=\tilde p=0$. But, this might describe massless ghost
states with negative norm that couple to a gauge field, leading to an
inconsistent theory \cite{peli}. We leave this discussion open in this
paper \footnote{I am grateful to N.B.~Nielsen and Y.~Shamir for discussions and
sharing their wisdom on this point.}. 

However, we would like to look at this point based on the point of view that is
the essential idea presented in the original paper of Eichten and
Preskill \cite{ep} to have chiral fermions in continuum limit. The question
of whether the spectra of normal modes of $\psi^i_L(x)$ and $\psi_R(x)$ are
massless and chiral is crucially related to the question of whether the normal
modes of the three-fermion states $(\bar\psi_R(x)\cdot\psi^j_L(x)) \psi_R(x)$
and $(\bar\psi^i_L(x)\cdot\psi_R(x)) \psi^i_L(x)$ have been composed in
segment {\it A}. As have been qualitatively discussed in section 5, the effective
multifermion coupling (the 1PI four-point vertex function (\ref{4p'})) is
strongly momentum dependent and thresholds and wedges for different modes
emerge in the phase diagram. This effective multifermion coupling for normal
modes is very small in segment {\it A}, thus preventing the normal modes of the
$\psi^i_L$ and $\psi_R$ from binding into three-fermion states
$(\bar\psi^i_L\cdot\psi_R)\psi^i_L$, $(\bar\psi_R\cdot\psi^i_L)\psi_R$. We can
reasonably speculate that there exists such a segment {\it A} with finite $g_2$, where
the scaling window of the continuum limit of chiral fermions would be opened
up, although we cannot give a rigorous demonstration for the time being. 

If what we expect is convincingly confirmed, the spectrum of the theory
in this segment will be the following. The spectrum consists of 15 copies of
$SU(2)$-QCD charged Dirac doublers ($p\not=0$) eq.~(\ref{sc11}) and 15 copies of $SU(2)$
neutral Dirac doublers ($p\not=0$) eq.~(\ref{sn11}). They are very massive and decoupled.
Besides, the low-energy spectrum contain the two massless Weyl modes
eqs.~(\ref{sb1},\ref{sb2}) for $g_1=0$ and $p=\tilde p$, 
\begin{equation} 
S^{-1}_L(\tilde p)^{ij}=i\gamma_\mu\tilde p^\mu\tilde Z_2\delta_{ij}P_L; 
\hskip1cm
S^{-1}_R(\tilde p)=i\gamma_\mu\tilde p^\mu P_R, 
\label{sf} 
\end{equation} which
is in agreement with the $SU_L(2)\otimes U_R(1)$ symmetry. Namely, this normal
mode of $\psi_L^i$ is self-scattering via the multifermion coupling $g_2$
(see Fig.~5) without pairing up with any other modes. The wave-function
renormalization $\tilde Z_2$ can be considered as an interpolating constant of
$Z_2(p)$ eq.~(\ref{z2}) for $p=\tilde p\simeq 0$ and $g_1=0$. In addition,
there is the $SU_L(2)\otimes U_R(1)$ covariant scalar ${\cal A}^i$
eq.~(\ref{mas}). In order to see all possible interactions between
these modes in this possible scaling region, we consider the one-particle
irreducible vertex functions of these modes. In the light of the exact
$SU_L(2)\otimes U_R(1)$ chiral symmetry and $\psi_R$-shift-symmetry, one can
straightforwardly obtain non-vanishing vertex functions ($d$=dimensions) at
physical momenta ($p=\tilde p, q=\tilde q$): (i) ${\cal A}^j{\cal
A}^{j\dagger}{\cal A}^i{\cal A}^{i\dagger}$ ($d=4$); (ii)
$\bar\psi^i_L\psi_L^i{\cal A}^j{\cal A}^{j\dagger}$, $\bar\Psi^i_c\Psi_c^i{\cal
A}^j{\cal A}^{j\dagger}$ and $\bar\Psi_n\Psi_n{\cal A}^j{\cal A}^{j\dagger}$
($d=5$), as well as $d>5$ vertex functions. The vertex functions with
dimensions $d>4$ vanish in the scaling region as $O(a^{d-4})$ and we are left
with the self-interacting vertex ${\cal A}^j{\cal A}^{j\dagger}{\cal A}^i{\cal
A}^{i\dagger}$. 

In this possible scaling region, the chiral continuum limit is very much like
that of lattice QCD. We need to tune only one coupling $g_1\rightarrow0$ in the
neighbourhood of the segment A $g_2^{c,a}<g_2<g_2^{c,\infty}$. For
$g_1\rightarrow 0$, the $\psi_R$ shift-symmetry is slightly violated, the
normal modes of $\psi^i_L$ and $\psi_R$ would couple together to form the
chiral symmetry breaking term $\Sigma^i(0)\bar\psi^i_L\psi_R$, which is a
dimension-3 renormalized operator and thus irrelevant at the short distance. We
desire this scaling region to be ultra-violet stable, in which the multifermion
coupling $g_1$ turns out to be an effective renormalized dimension-4
operator \cite{bar}. 

\section{Some remarks}

The conclusion of the existence of the possible scaling region (\ref{segment})
for the continuum chiral theory is plausible and hard to be excluded. It is
worthwhile to check this scenario in different approaches. Even though this
scenario emerges, we are still left with several other problems to have a
correct continuum chiral gauge theory when the global $SU_L(2)$ chiral symmetry
is gauged. Their possible resolutions are mentioned and discussed in this
section, and deserve to be studied in future work. 

The question is whether this chiral continuum theory in the scaling region
could be the correct chiral gauge theory, as the $SU(2)$ chiral gauge coupling
$g$ is perturbatively turned on in the theory (\ref{action}). One should expect
a slight change of critical lines (points). We should be able to re-tune the
multifermion couplings ($g_1,g_2$) to compensate these perturbative changes, due
the fact that the gauge interaction does not spoil the $\psi_R$ shift-symmetry
and we can derive identities based on the Ward identity (\ref{w}),
\begin{equation}
{\delta^{(2)} \Gamma\over\delta
A'_\mu\delta\bar\psi'_R}={\delta^{(3)} \Gamma\over\delta
A'_\mu\delta\psi'_R\delta\bar\psi'_R}={\delta^{(3)} \Gamma\over\delta
A'_\mu\delta\Psi'^n_L\delta\bar\psi'_R}=\cdot\cdot\cdot=0,
\nonumber
\end{equation}
where $A'_\mu$ is a ``prime'' gauge field.
In this possible scaling regime, disregarding those uninteresting neutral
modes, we have the charged modes including both the $SU(2)$ chiral-gauged,
massless normal mode (\ref{sf}) of $\psi^i_L$ and the
fifteen $SU(2)$-vectorial-gauged, massive doublers of the Dirac fermion 
$\Psi^i_c$ (\ref{sc11}). The gauge
field should not only chirally couple to the massless normal mode of the
$\psi_L^i$ in the low-energy regime, but also vectorially couple to the massive
doublers of Dirac fermion $\Psi_c^i$ in the high-energy regime. Thus, we expect
the coupling vertex of the $SU_L(2)$-gauge field and the normal mode of the
$\psi^i_L$ to be chiral at the continuum limit. We should be able to
demonstrate this point on the basis of the Ward identities associated with the
$SU(2)$ chiral gauge symmetry that is respected by the spectrum in the possible
scaling regime. In fact, due to the reinstatement of the manifest
$SU_L(2)$ chiral gauge symmetry and corresponding Ward identities of the
undoubled spectrum in this possible scaling regime, we should then apply the
Rome approach \cite{rome} (which is based on the conventional wisdom of quantum
field theories) to perturbation theory in the small gauge coupling. It is
expected that the Rome approach would work in the same way but all
gauge-variant counterterms are prohibited; the gauge boson masses vanish to all
orders of gauge coupling perturbation theory for $g_1=0$. We hope not to 
run into an inconsistent theory if the propagators of the three-fermion states 
which are composed of the normal modes of $\psi_L^i$ and $\psi_R$, 
positively vanish when $p\rightarrow 0$, i.e., the zero of these 
propagators at
$p=0$ is a double zero. 

Another important question remaining is how chiral gauge anomalies emerge,
although in this short report the chiral gauge anomaly is cancelled by
purposely choosing an appropriate fermion representation of the $SU_L(2)$
chiral gauge group. We know that in the doubled spectrum of naive lattice
chiral gauge theory, the reason for the correct anomaly disappearing in the
continuum limit is that the normal mode and doublers of Weyl fermions produce
the same anomaly , so that these anomalies eliminate themselves \cite{smit}. As a
consequence of decoupled doublers being given a chiral-invariant mass $(\sim
O({1\over a}))$, the surviving normal mode of the Weyl fermion (chiral-gauged,
e.g., $U_L(1)$) should produce the correct anomaly in the continuum limit. Due
to the fact that the action (\ref{action}) possesses explicit $U_L(1)$ global
symmetry, we also have the question of whether the conservation of left-handed
fermion number would be violated by the correct anomaly \cite{ep,bank} structure
$\tr F\tilde F$ that is generated by the $SU(2)$ instanton in the continuum
limit. We will discuses this crucial problem in a separate paper. 

I thank G.~Preparata, M.~Creutz and H.B.~Nielsen for many discussions. 
Thanks to R.~Shrock and M.~Testa for discussions on multifermion
couplings, and P.G.~Ratcliffe for reading this paper. The author gratefully
acknowledges the support of K.C.~Kong education of foundation, Hong Kong
and the national foundation of science, China.

\vskip1.5cm
\appendix\noindent {\bf Appendix I}
\vskip1cm
To obtain eq.~(\ref{wf}), we need to consider
\begin{eqnarray}
\langle\bar\psi^{i\alpha}_L(x)\psi^\alpha_R(x)\psi_L^{i\beta}(x)\rangle
&\equiv &-{1\over Z}{\delta\over \delta\eta^{i\alpha}_L(x)}
{\delta\over \delta\bar\eta^\alpha_R(x)}{\delta\over \delta\bar\eta^{i\beta}_L
(x)} Z\nonumber\\
&=&{1\over Z}{\delta\over \delta\eta^{i\alpha}_L(x)}
{\delta\over \delta\bar\eta^{i\beta}_L(x)}\left[ Z{1\over Z}
{\delta\over \delta\bar\eta^\alpha_R(x)} Z\right]\nonumber\\
&=&{1\over Z}{\delta\over \delta\eta^{i\alpha}_L(x)}
{\delta\over \delta\bar\eta^{i\beta}_L(x)}
\left[ Z(-)\psi'^\alpha_R(x) \right],
\label{d}
\end{eqnarray}
where we utilize the relation (\ref{pr}), then
\begin{eqnarray}
{\delta\over\delta\psi'^\alpha_R(0)}
\langle\bar\psi^{i\sigma}_L(x)\psi^\sigma_R(x)\psi_L^{i\beta}(x)\rangle
&=&-{1\over Z}{\delta\over \delta\eta^{i\alpha}_L(x)}
{\delta\over \delta\bar\eta^{i\beta}_L(x)}\left[ Z\delta(x)\right],\nonumber\\
&\equiv&-\langle\bar\psi^{i\alpha}_L(x)\psi_L^{i\beta}(x)\rangle\delta(x).
\label{der1}
\end{eqnarray}
Analogously, one can have,
\begin{equation}
{\delta\over\delta\psi'^\alpha_R(0)}
\langle\Delta\left(\bar\psi^{i\sigma}_L(x)\Delta(\psi^\sigma_R(x))
\psi_L^{i\beta}(x)\right)\rangle
=-\Delta\left(\langle\bar\psi^{i\alpha}_L(x)\psi_L^{i\beta}(x)\rangle
\Delta\delta(x)\right)\label{der2}
\end{equation}
eqs.~(\ref{der1},\ref{der2}) lead to eq.~(\ref{wf}).
 
$\langle\bar\psi^{i\alpha}_L(x)\psi_L^{i\beta}(y)\rangle_\circ$ is a
spinor matrix that can be expanded on the basis of $I,\gamma_\mu, 
\sigma_{\mu\nu}, \gamma_5\gamma_\mu$ and $\gamma_5$, non-vanishing terms
can be written as
\begin{eqnarray}
\langle\bar\psi^{i\alpha}_L(x)\psi_L^{i\beta}(y)\rangle_\circ
&=&(P_L)^{\alpha\sigma}
\langle\bar\psi^{i\sigma}(x)\psi^{i\theta}(y)\rangle_\circ (P_R)^{\theta\beta}
\label{pt1}\\
&=&(P_L)^{\alpha\sigma}
\int_pe^{ip(x-y)}
\left(A(\bar p^2)
(\gamma_\mu)^{\sigma\theta}\bar p_\mu+B(\bar p^2)
(\gamma_5\gamma_\mu)^{\sigma\theta}\bar p_\mu \right)(P_R)^{\theta\beta},
\nonumber
\end{eqnarray}
where $\int_p=\int_{-\pi}^\pi{d^4p\over (2\pi)^4}$ and 
$\bar p_\mu=\sin (p_\mu)$ is the lattice momentum. For $x=y$,
one gets eq.~(\ref{check}),
\begin{equation}
\langle\bar\psi^{i\alpha}_L(x)\psi^{i\beta}_L(x)\rangle_\circ=0,\hskip2cm
\Delta\langle\bar\psi^{i\alpha}_L(x)\psi^{i\beta}_L(x)\rangle_\circ=0.
\label{pt2}
\end{equation}
because of an odd function integral.

To obtain eq.~(\ref{ws1}), analogously to eqs.~(\ref{d},\ref{der1},\ref{der2}), we 
need to calculate 
\begin{eqnarray}
\langle\bar\psi^{i\sigma}_L(x)\psi^\sigma_R(x)\psi_L^{i\beta}(x)\rangle
&=&{1\over Z}{\delta\over \delta\eta^{i\sigma}_L(x)}
{\delta\over \delta\bar\eta^\sigma_R(x)}
\left[ Z\psi'^{i\beta}_L(x) \right],\nonumber\\
{\delta\over\delta\psi'^{i\alpha}_L(0)}
\langle\bar\psi^{i\sigma}_L(x)\psi^\sigma_R(x)\psi_L^{i\beta}(x)\rangle
&=&\langle\bar\psi^{i\sigma}_L(x)\psi_R^\sigma (x)\rangle \delta(x)
\delta_{\alpha\beta},
\label{der3}
\end{eqnarray}
and 
\begin{equation}
{\delta\over\delta\psi'^{i\alpha}_L(0)}
\langle\Delta\left(\bar\psi^{i\sigma}_L(x)\Delta(\psi^\sigma_R(x))
\psi_L^{i\beta}(x)\right)\rangle
=-\Delta\left(\langle\bar\psi^{i\sigma}_L(x)\psi_R^\sigma(x)\rangle
\Delta\delta(x)\right)\delta_{\alpha\beta}.
\label{der4}
\end{equation}
Armed with these equations, we obtain eq.~(\ref{ws1}) in section 3.

\vskip1.5cm
\noindent
\appendix {\bf Appendix II}
\vskip1cm
We re-insert the parametrization of (\ref{para}) into the RHS of
gap-equation (\ref{se}), and equaling this to eq.~(\ref{para}) gives
us two equations 
\begin{eqnarray}
\rho &=& 4\tilde g_1\int_q{(\tilde g_2w(q)+\rho)\over 
\sum_\rho\sin^2q_\rho +(\rho+\tilde g_2w(q))^2(v^ia)^2
}\nonumber\\
1 &=& 4\int_q{(\tilde g_2w(q)+\rho)\over 
\sum_\rho\sin^2q_\rho +(\rho+\tilde g_2w(q))^2(v^ia)^2
}w(q).\label{drho1}
\end{eqnarray}
With the definition $I_n$ (\ref{in}) for $n=0,1,2$, we have equations
\begin{eqnarray}
\rho &=& \tilde g_1\tilde g_2I_1+\tilde g_1\rho I_\circ\nonumber\\
1 &=& 4\tilde g_2I_2+4\rho I_1,\label{drho}
\end{eqnarray}
which lead to eq.~(\ref{rho}).

The free propagator of $\psi_L^i$ is given as
\begin{equation}
S_\circ^{ij}(p)=\delta_{ij}P_L\hat pP_R,\hskip1cm \hat p={{i\over a}\sum_\mu\gamma^\mu
\sin p_\mu\over {1\over a^2}\sum_\mu\sin^2p_\mu}a^2.
\nonumber
\end{equation}
The Feynman diagram (see Fig.~4) is given by,
\begin{eqnarray}
\hat\sigma^{ij}(p)&=&\delta^{ij}P_R\sigma(p)P_L\nonumber\\
\sigma^{ij}(p)&=&-a^{-2}\int_k{{i\over a}\sum_\mu\gamma^\mu
\sin(p-k)_\mu\over {1\over a^2} \sum_\mu\sin^2(p-k)_\mu}\nonumber\\
&&\cdot\int_q\tr\left[
P_L{{i\over a}\sum_\mu\gamma^\mu
\sin(k-{q\over2})_\mu\over  {1\over a^2} \sum_\mu\sin^2(k-{q\over2})_\mu}P_R
{{i\over a}\sum_\mu\gamma^\mu
\sin(k+{q\over2})_\mu\over
{1\over a^2}\sum_\mu\sin^2(k+{q\over2})_\mu}P_L\right]\lambda 
\nonumber\\
&&=2aN_f\int_{k,q}\lambda {i\sum_\mu\gamma^\mu
\sin(p-k)_\mu\over \sum_\mu\sin^2(p-k)_\mu} R(k,q),
\label{sigma}
\end{eqnarray}
where $R(k,q)$ is eq.~(\ref{z2}) and 
\begin{equation}
\lambda=\left(g_1+4g_2w(p-k)w(k+{q\over2})\right)^2
={1\over N_f^2}a^{-4}\left(\tilde g_1+4\tilde g_2w(p-k)w(k+{q\over2})\right)^2.
\nonumber
\end{equation}

The wave-function renormalization $Z_2(p)$ of $\psi^i_L(x)$ in
eq.~(\ref{wave}) can be calculated by using the train approximation (see Fig.~5),
\begin{eqnarray}
Z_2^{-1}S_\circ^{ij}(p)&=&P_L(\hat p+\hat p\sigma\hat p+\hat p\sigma\hat p
\sigma\hat p+\cdot\cdot\cdot)P_R\delta^{ij}
\nonumber\\
&=&S_\circ^{ij}(p)\left({1\over 1-\sigma\hat p}\right),
\label{train}
\end{eqnarray}
and one gets 
\begin{equation}
Z_2=1-\sigma\hat p.
\label{az2}
\end{equation}
With eq.~(\ref{sigma}) one can get 
\begin{equation}
\sigma\hat p=-
{2\over N_f}\!\int_{k,q}\!\left(\tilde g_1\!+\!4
\tilde g_2w(p-k)w(k\!+\!{q\over2})\right)^2\!
{\sum_{\mu\nu}\!\gamma_\mu\gamma_\nu
\sin(p\!-\!k)^\mu\sin p^\nu\over\sum_{\lambda\rho}\sin^2(p-k)_\lambda
\sin^2 p_\rho}R(k,q).
\label{az1}
\end{equation}
By substituting eq.~(\ref{az1}) into (\ref{az2}), one gets eq.~(\ref{z2}).

\vskip1.5cm
\noindent
\appendix {\bf Appendix III}
\vskip1cm
We keep in mind that 
$\psi_L^i$ has two components and $\psi_R$ has one component (all times 
a factor of 2 for spin degeneracy). In the strong-coupling limit, the
kinetic terms $S_f(x)$:
\begin{eqnarray}
S_f(x)&=&S_f^L(x)+S_f^R(x)\nonumber\\
S_f^L(x)&=&{1\over 2ag_2^{1\over2}}\sum_\mu\bar\psi^i_L(x)
\gamma_\mu\partial^\mu\psi^i_L(x)
;\hskip0.5cm S_f^R(x)={1\over 2ag_2^{1\over2}}
\sum_\mu\bar\psi_R(x)\gamma_\mu\partial^\mu\psi_R(x)
\label{kk}
\end{eqnarray}
are dropped, the interacting
action $S_2(x)$ (\ref{rs21}) turns out to be bilinear in $\psi^i_L(x)$ fermion field 
at the same point ``$x$''.
We first perform the integral of $\psi_L^i(x)$ ($i$ fixed) at the point ``$x$''and 
obtain
\begin{eqnarray}
D^i(x)&\equiv &\Pi_\alpha\int[d\bar\psi_L^{i\alpha} d\psi_L^{i\alpha}]\exp\left(-
\bar\psi^i_L(x)\cdot\left[\Delta\psi_R(x)\right]
\left[\Delta\bar\psi_R(x)\right]\cdot\psi_L^i(x)\right)\nonumber\\
&=&\det_s\left(\matrix{ \Delta\bar\psi_R^1(x)\Delta\psi_R^1(x)&
\Delta\bar\psi_R^1(x)\Delta\psi_R^2(x)\cr
\Delta\bar\psi_R^2(x)\Delta\psi_R^1(x)&
\Delta\bar\psi_R^2(x)\Delta\psi_R^2(x)\cr}\right)\nonumber\\
&=&2\Pi_\alpha\Delta\psi^\alpha_R(x)\Delta\bar\psi^\alpha_R(x),
\label{detl}
\end{eqnarray}
where the determent $\det_s$ is taken over spinor space. Then, the partition
function of the one-site theory is given by the integral that 
is bilinear in fermionic variable $\psi^{\alpha}_R(x)$,
\begin{eqnarray}
Z^i(x)&=&\Pi_\alpha\int[d\bar\psi_R^\alpha(x) d\psi_R^\alpha(x)]D^i(x)\nonumber\\
&=&\Pi_\alpha\left(\int[d\bar\psi_R^\alpha(x) d\psi_R^\alpha(x)]
2\Delta\psi^\alpha_R(x)\Delta\bar\psi^\alpha_R(x)\right)\nonumber\\
&=&2^2\det_s\left(\Delta^2(x)\right)=2^2\left(\Delta^2(x)\right)^2.
\label{az}
\end{eqnarray}
The total partition function in the strong-coupling limit is then obtained,
\begin{eqnarray}
Z&=&\Pi_{i,x}\left(\Pi_\alpha\int[d\bar\psi_R^\alpha(x) d\psi_R^\alpha(x)]
2\Delta\psi^\alpha_R(x)\Delta\bar\psi^\alpha_R(x)\right)\nonumber\\
&=&2^{4N}\left(\det_x\Delta^2(x)\right)^4,
\label{stronglimit'}
\end{eqnarray}
where the determent $\det_x$ is taken only over the lattice space-time and 
$N$ is the number of lattice sites. 

In order to obtain the recursion relations (\ref{re1},\ref{re2},\ref{re3}),
(\ref{re11},\ref{re21},\ref{re31}) and (\ref{re4}) satisfied by two-point
functions, we consider an integral of one field $\psi^j_L(x)$ at the point
``$x$'' defined as 
\begin{equation}
P^{j\sigma}_1(x)\equiv {1\over Z^j(x)}\int_x^R\int_{xj}^L
\bar\psi^{j\sigma}_L(x)e^{-S_f(x)-S^j_2(x)},
\label{p1}
\end{equation}
where the measure of the fermion fields $\psi^j_L(x)$ 
(the $SU_L(2)$-index $j$ is fixed) and $\psi_R(x)$ at the point 
``$x$'' is given as
\begin{equation}
\int_x^R\int_{xj}^L\equiv\Pi_\alpha\int 
[d\bar\psi_R^\alpha(x)d\psi_R^\alpha(x)]
[d\bar\psi_L^{j\alpha}(x)d\psi_L^{j\alpha}(x)].
\nonumber
\end{equation}
To have
a non-vanishing integral of $\psi_L^j(x)$ at the point ``$x$'', we need
a $\psi_L^j(x)$ field in the expansion of $e^{-S^L_f(x)}$, and obtain
\begin{equation}
P^{j\sigma}_1(x)\!=\!{1\over Z^j(x)}\left({1\over 2a
g_2^{1\over2}}\right)\sum^\dagger_\mu[\bar\psi^j_L(x\!+\!\mu)\gamma_\mu]^\gamma
\int_x^R\int_{xj}^L\psi^{j\gamma}_L(x)
\bar\psi^{j\sigma}_L(x)e^{-S_f^R(x)-S^j_2(x)},\label{pp}
\end{equation}
where
\begin{equation}
\sum^\dagger_\mu[\bar\psi^j_L(x\!+\!\mu)\gamma_\mu]^\gamma
=[\bar\psi^j_L(x\!+\!\mu)\gamma_\mu]^\gamma-
[\bar\psi^j_L(x\!-\!\mu)\gamma_\mu]^\gamma.\nonumber
\end{equation}
Using eq.~(\ref{detl}), we can first perform the integral over 
$\psi_L^j(x)$ in eq.~(\ref{pp}) that is bilinear in terms of $\psi_L^j(x)$, 
\begin{eqnarray}
\int_{xj}^L\psi^{j\gamma}_L(x)
\bar\psi^{j\sigma}_L(x)e^{-S^j_2(x)}
&\!=\!&
\left({D^j(x)\over\Delta\psi^\gamma_R(x)\Delta\bar\psi^\sigma_R(x)}\right)
\label{intl}\\
\left({D^j(x)\over\Delta\psi^\gamma_R(x)\Delta\bar\psi^\sigma_R(x)}\right)
&\!=\!&-\delta_{\gamma 1}\delta_{\sigma 1}\Delta\psi^2_R(x)\Delta\bar\psi^2_R(x)
-\delta_{\gamma 2}\delta_{\sigma 2}\Delta\psi^1_R(x)\Delta\bar\psi^1_R(x)
\nonumber\\
&\!+\!&\delta_{\gamma 1}\delta_{\sigma 2}\Delta\psi^2_R(x)\Delta\bar\psi^1_R(x)
+\delta_{\gamma 2}\delta_{\sigma 1}\Delta\psi^1_R(x)\Delta\bar\psi^2_R(x).
\nonumber
\end{eqnarray}
As a result, we have
\begin{equation}
P^{j\sigma}_1(x)=\left({1\over 2a
g_2^{1\over2}}\right)\sum^\dagger_\mu[\bar\psi^j_L(x\!+\!\mu)\gamma_\mu]^\gamma
{1\over Z^j(x)}
\int_x^R e^{-S_f^R(x)}
\left({D^j(x)\over\Delta\psi^\gamma_R(x)\Delta\bar\psi^\sigma_R(x)}\right).
\label{l1}
\end{equation}
The remaining integral (\ref{l1}) of $\psi_R(x)$ at the ``$x$'' point 
is bilinear in terms of fermionic variable $\psi^\alpha_R(x)$, we need to have
$\psi_R(x)$ and $\bar\psi_R(x)$ fields in expansion of $e^{-S_f^R(x)}$, 
for the lowest order of $O({1\over g_2})$, we obtain 
\begin{equation}
P^{j\sigma}_1(x)=\left({1\over 2a
g_2^{1\over2}}\right)^3\sum^\dagger_\mu [\bar\psi^j_L(x+\mu)\gamma_\mu]^\gamma
[\gamma_\mu\psi_R(x+\mu)]^\alpha[\bar\psi_R(x+\mu)\gamma_\mu]^\beta
\Theta^j_{\beta\alpha\gamma\sigma}(x),\nonumber
\end{equation}
where
\begin{equation}
\Theta^j_{\beta\alpha\gamma\sigma}(x)={1\over Z^j(x)}
\Pi_\lambda\int [d\bar\psi_R^\lambda(x)d\psi_R^\lambda(x)]\psi_R^\beta(x)
\bar\psi^\alpha_R(x)
\left({D^j(x)\over\Delta\psi^\gamma_R(x)\Delta\bar\psi^\sigma_R(x)}\right).
\label{p1m}
\end{equation}
Using eqs.~(\ref{az},\ref{intl}) and the following relation
\begin{eqnarray}
&&{1\over Z^j(x)}\Pi_\lambda\int [d\bar\psi_R^\lambda(x)d\psi_R^\lambda(x)]
\left[\Delta\psi^\gamma_R(x)\Delta\bar\psi^\sigma_R(x)\right]
\psi_R^\beta(x)\bar\psi^\alpha_R(x)\nonumber\\
&=&{1\over Z^j(x)}{\delta_{\beta\sigma}\delta_{\gamma\alpha}\over \Delta^2(x)}
2^2\det_s\left(\Delta^2(x)\right)=
{\delta_{\beta\sigma}\delta_{\gamma\alpha}\over \Delta^2(x)},
\label{indelta}
\end{eqnarray}
we obtain
\begin{eqnarray}
\Theta^j_{\beta\alpha\gamma\sigma}&=&{\delta_{\beta\sigma}\delta_{\gamma\alpha}
\over \Delta^2(x)}\label{theta}\\
P^{j\sigma}_1(x)&=&{1\over \Delta^2(x)}\left({1\over 2a
g_2^{1\over2}}\right)^3\sum^\dagger_\mu [\bar\psi^j_L(x\!+\!\mu)
\cdot\psi_R(x\!+\!\mu)][\bar\psi_R(x\!+\!\mu)\gamma_\mu]^\sigma.
\label{p1'}
\end{eqnarray}
In eq.~(\ref{p1'}) the reason for the three fields $\psi_R,\bar\psi_R$ and 
$\psi_L^i$ being at the same point ``$x+\mu$'' is due to the lowest 
non-trivial approximation.

We define the integral of three fields $P^{j\sigma}_3(x)$ at the site ``$x$''
\begin{equation}
P^{j\sigma}_3(x)\equiv {1\over Z^j(x)}\int_x^R\int_{xj}^L
\bar\psi^{j\alpha}_L(x)\psi^\alpha_R(x)
\bar\psi^\sigma_R(x)e^{-S_f(x)-S^j_2(x)}.
\label{p3}
\end{equation}
Analogously to the reason for eqs.~(\ref{l1},\ref{p1m}) to get non-trivial
results, we only need a $\psi_L^j(x)$ field in the expansion of
$e^{-S^L_f(x)}$, and considering eqs.~(\ref{intl},\ref{theta}) we obtain, 
\begin{eqnarray}
P^{j\sigma}_3(x)\!&=&\!{1\over Z^j(x)}\left({1\over 2a
g_2^{1\over2}}\right)\sum^\dagger_\mu[\bar\psi^j_L(x\!+\!\mu)\gamma_\mu]^\gamma
\int_x^R\int_{xj}^L\psi^{j\gamma}_L(x)\bar\psi^{j\alpha}_L(x)\psi^\alpha_R(x)
\bar\psi^\sigma_R(x)
e^{-S^j_2(x)}\nonumber\\
&=&\left({1\over 2a
g_2^{1\over2}}\right)\sum^\dagger_\mu[\bar\psi^j_L(x\!+\!\mu)\gamma_\mu]^\gamma
\Theta^j_{\beta\sigma\gamma\beta}(x),\nonumber\\
&=&{1\over \Delta^2(x)}\left({1\over 2a
g_2^{1\over2}}\right)\sum^\dagger_\mu [\bar\psi^j_L(x+\mu)
\gamma_\mu]^\sigma.
\label{p3'}
\end{eqnarray}
We turn to consider the integral of the four fermion fields at site ``$x$'',
\begin{eqnarray}
P^{ij,\theta\sigma}_4(x)&\equiv& {1\over Z^j(x)}\int_x^R\int_{xj}^L
\psi^{\theta i}_L(0)\bar\psi^{j\alpha}_L(x)\psi^\alpha_R(x)
\bar\psi^\sigma_R(x)e^{-S_f(x)-S^j_2(x)}\nonumber\\
\!&\simeq&\!{1\over Z^j(x)}
\int_x^R\int_{xj}^L\psi^{i\theta}_L(0)\bar\psi^{j\alpha}_L(x)\psi^\alpha_R(x)
\bar\psi^\sigma_R(x)
e^{-S^j_2(x)}\nonumber\\
&=&\delta(x)\delta_{ij}\Theta_{\alpha\sigma\theta\alpha}(x)
\nonumber\\
&=&{\delta_{\theta\sigma}\delta (x)\delta_{ij}\over \Delta^2(x)}.
\label{p4}
\end{eqnarray}
We compute the following integral $P^j_2$(x) of two fields 
$\bar\psi^{j\alpha}_L(x)\psi^\alpha_R(x)$ at ``$x$'',
\begin{equation}
P^j_2(x)\equiv {1\over Z^j(x)}\int_x^R\int_{xj}^L
\bar\psi^{j\alpha}_L(x)\psi^\alpha_R(x)e^{-S_f(x)-S^j_2(x)}.
\label{p2}
\end{equation}
To have a non-trivial result, we need
$\psi_L^j(x)$ and $\psi_R(x)$ fields in the expansion of $e^{-S_f(x)}$, 
and we obtain
\begin{eqnarray}
P^j_2(x)\!&=&\!{1\over Z^j(x)}\left({1\over 2a
g_2^{1\over2}}\right)\sum^\dagger_\mu[\bar\psi^i_L(x\!+\!\mu)\gamma_\mu]^\gamma
\int_x^R\int_{xj}^L\psi^{j\gamma}_L(x)\bar\psi^{j\alpha}_L(x)\psi^\alpha_R(x)
e^{-S_f^R(x)-S^j_2(x)}\nonumber\\
&=&\left({1\over 2a
g_2^{1\over2}}\right)^2
\sum_{\pm\mu}[\bar\psi^j_L(x\!+\!\mu)\gamma_\mu]^\gamma
[\gamma_\mu\psi_R(x\!+\!\mu)]^\sigma\Theta_{\alpha\sigma\gamma\alpha}(x)
\nonumber\\
&=&{1\over \Delta^2(x)}
\left({1\over 2a
g_2^{1\over2}}\right)^2
\sum_{\pm\mu}[\bar\psi^j_L(x\!+\!\mu)\psi_R(x\!+\!\mu)].
\label{p2'}
\end{eqnarray}
Armed with these equations (\ref{p1'},\ref{p3'},\ref{p4},\ref{p2'}), 
one can get the recursion relations satisfied by two-point functions in the
main text of the article.

\section*{Figure Captions} 
 
\noindent {\bf Figure 1}: \hspace*{0.2cm} 
The phase diagram for the theory (\ref{action}) in the $g_1-g_2$ plane.

\noindent {\bf Figure 2}: \hspace*{0.2cm} 
The effective four-point interacting vertex.

\noindent {\bf Figure 3}: \hspace*{0.2cm} 
The NJL gap-equation for the self-energy function $\Sigma(p)$ in the limit
of $N_f\rightarrow\infty$.

\noindent {\bf Figure 4}: \hspace*{0.2cm} 
The Feynman diagram of the leading contribution to the wave-function 
renormalization $Z_2(p)$ of $\psi_L^i$.

\noindent {\bf Figure 5}: \hspace*{0.2cm} 
The train approximation to the wave-function renormalization $Z_2(p)$ of 
$\psi_L^i$.

\end{document}